\documentclass[%
reprint,
frontmatterverbose, 
nofootinbib,
amsmath,amssymb,
aps,showkeys,showpacs,
prd,
prstab,
prstper,
]{revtex4-1}
\usepackage{newtxmath,newtxtext}
\usepackage{graphicx}
\usepackage[caption=false]{subfig}
\usepackage{dcolumn}
\usepackage{bm}
\usepackage{dsfont}
\usepackage{amsmath,amssymb,stackengine}
\usepackage{siunitx}
\usepackage[caption=false]{subfig}
\usepackage{braket}
\usepackage{listings}
\usepackage{xcolor}
\usepackage{hyperref}

\usepackage{tcolorbox}
\usepackage{braket}
\hypersetup{colorlinks=true,linkcolor=blue,urlcolor=blue,citecolor=blue}
\usepackage{graphicx}
\usepackage{dcolumn}
\usepackage{bm}

\usepackage{orcidlink}
\usepackage{multirow, makecell}
\usepackage{rotating}
\usepackage{tikz}

\begin{document}
\title{Reconciling fractional entropy and black hole entropy compositions}

\author{Manosh T. Manoharan\orcidlink{0000-0001-8455-6951}}
\email{tm.manosh@gmail.com; tm.manosh@cusat.ac.in}
\author{N. Shaji\orcidlink{0000-0002-5669-625X}}
\email{shajin@cusat.ac.in}
\affiliation{Department of Physics, Cochin University of Science and Technology, Kochi -- 682022, India.}%


\begin{abstract}
This study investigates the implications of adopting fractional entropy in the area law framework and demonstrates its natural alignment with an isothermal description of black hole composition. We discuss the Zeroth law compatibility of the fractional entropy and define an empirical temperature for the horizon. We highlight the distinction between the empirical and conventional Hawking temperatures associated with the black holes. Unlike the Hawking temperature, this empirical temperature appears universal, and its proximity to the Planck temperature suggests a possible quantum gravity origin. We also establish the connection between these temperatures. Furthermore, extending the conventional fractional parameter $q$, constrained between 0 and 1, we establish that any positive real number can bound $q$ under the concavity condition, provided the log of micro-state dimensionality exceeds $q-1$. Specifically, for black holes, $q = 2$, necessitating micro-state dimensionality greater than $e$, thereby excluding the construction of black hole horizon states with two level bits or qubits. We also identify the connection between the validity of the second law and information fluctuation complexity. The second law requires that the variance of information content remain smaller than the area of the black hole horizon. This constraint naturally gives rise to a Boltzmann-Gibbs-like entropy for the black hole, which, in contrast to the canonical formulation, is associated with its mass rather than its area. Equilibrium distribution analysis uncovers multiple configurations, in which the one satisfying the prerequisites of probability distribution exhibits an exponent stretched form, revealing apparent deviation from the Boltzmann distribution. 
\end{abstract}

\maketitle

\section{Introduction}

The connection between thermodynamics and gravity has been a subject of attraction since the discovery of the Fulling-Davies-Unruh effect \cite{PhysRevD.7.2850, PCWDavies_1975,PhysRevD.14.870}. This relationship was further developed into a significant area of study through the formulation of horizon mechanics \cite{Bardeen1973, PhysRevD.15.2738,PhysRevD.23.287,PhysRevD.7.2333,PhysRevD.9.3292,PhysRevD.13.191,Hawking1975,HAWKING1974}. It is evident that gravitational entropy, as expressed by the area law, is associated specifically with the horizon rather than traditional statistical thermodynamic entropy \cite{PhysRevD.15.2752}. This area law applies to cosmological and black hole horizons, collectively called the laws of horizon thermodynamics.

An often-overlooked aspect of horizon thermodynamics is the notion of thermal equilibrium and its statistical origin. Thermal equilibrium is defined within a single horizon rather than between multiple horizons. For a horizon in equilibrium, the temperature remains uniform across the horizon and is proportional to the associated surface gravity. The energy considered is the total energy that constitutes the horizon. This energy is not additive in the conventional sense, as the horizon is a global feature of spacetime, distinguishing it from equilibrium thermodynamics in the standard sense.

Furthermore, the laws of horizon thermodynamics are well established for static and dynamic horizons \cite{PhysRevD.48.R3427,PhysRevD.50.846}. In general relativity, gravitational entropy is connected to the Noether charge, with first-order corrections expressed through the area law at the apparent horizon \cite{PhysRevD.110.024070}. This framework underpins the physical process version of the first law of thermodynamics \cite{PhysRevD.64.084020}, which states that any process eventually leads to a new stationary state where changes in entropy are related to variations in other observable parameters. The generalized second law also asserts that only processes increasing total entropy are allowed. This principle has been applied to astrophysical events like binary black hole mergers. For instance, Isi et al. \cite{PhysRevLett.127.011103} demonstrated that for the gravitational wave event GW150914, the entropy of the final black hole exceeded the combined entropies of the two inspiraling black holes.

In conventional horizon mechanics, black hole mergers are inherently non-isothermal processes. The final black hole typically has a lower temperature than the initial black hole. The laws of horizon mechanics, which assume isothermal conditions, only apply to individual black holes when they are well-separated initially or when the remnant is stable after the merger. However, the Boltzmann-Gibbs framework cannot describe the black hole composition process.

Let us examine the addition of entropy in black hole composition processes. The entropy $S$ of a Kerr black hole with angular momentum $J$ and mass $M$ is given by,
\begin{equation}
S(M, J) = k_B\left(\frac{4\pi r_h^2}{4\ell_p^2}\right),
\label{eq:KerrEntropy}
\end{equation}  
where $r_h = GM/c^2 + \sqrt{(GM/c^2)^2 - (J/Mc)^2}$, $\ell_p$ is the Planck length, $G$ is the universal gravitational constant, and $c$ is the causal speed limit. If each black hole has an entropy corresponding to the above expression. The total initial entropy is,
\begin{equation}
S_0 = S(M_1, J_1) + S(M_2, J_2).
\end{equation}  
After the merger, when the final black hole remnant reaches equilibrium, we have
\begin{equation}
S_f = S(M_f, J_f).
\end{equation}  
Although the physical processes involved in black hole mergers are highly complex, observations of binary black hole mergers suggest that the energy carried away by gravitational waves is small compared to the total mass. As a result, ignoring the entropy contributions from dissipation does not violate the generalized second law. Specifically, even though $M_f \lesssim M_1 + M_2$, the final entropy $S_f > S_0$ in all practical scenarios \cite{GWOSC_GWTC2024}. This indicates that entropy is produced during the merger process, and the final entropy must exceed the initial total entropy $S_0$ \cite{ThomasWKephart_2003}.  

Furthermore, for the idealized case of colliding Schwarzschild black holes, the maximum energy emission has an upper bound of approximately $\sim29\%$ \cite{PhysRevLett.26.1344}. While including spin and relativistic effects adds complexity, we focus here on an ideal process. For Schwarzschild black holes with negligible emission, energy conservation leads to 
\begin{equation}
S_f = S(M_1) + S(M_2) + 2\sqrt{S(M_1)S(M_2)}.
\end{equation}

This manuscript addresses several unresolved questions associated with the above description through the lens of statistical mechanics. In the following sections, we detail the conventional framework of horizon thermodynamics and establish the specific questions we aim to investigate within the context of non-additive entropies. Subsequently, we construct a fractional entropy tailored to address the problem. We examine the general features of this construction, extending it beyond its formal boundaries, and establish a robust connection between fractional entropy composition and black hole composition. This connection reveals a non-trivial relationship with Boltzmannian state counting. For the first time, we demonstrate the implications of this connection in the context of information fluctuation complexity, providing additional bounds on the validity of the second law of thermodynamics. Furthermore, we explore the empirical temperature compatible with the Zeroth law and its implications for understanding the microscopic degrees of freedom associated with the horizon. Our analysis suggests an additional discretization of mass based on fractional entropy, complementing the area discretization derived from conventional Boltzmannian state counting. We also investigate the standard thermodynamic and statistical features associated with this new entropy. We examine the equilibrium thermal distribution, identifying physically viable solutions and analyzing their properties, such as thermal stability. Our findings suggest the emergence of fractional dimensions and highlight the limitations of conventional bits or qubits in characterizing systems like black hole horizons in the context of composition processes.

\section{Area law, Additivity and Entropy Composition}

In the standard framework of black hole thermodynamics, the Hawking temperature of a black hole is defined by the relation,
\begin{align}
\frac{1}{T} = \frac{dS}{dE},
\end{align}
where $ T $ is the temperature, $ S $ is the entropy, and $ E $ is the energy. Classically, this can be derived using the area law, which relates the black hole entropy to the area of its event horizon. While both temperature and entropy have quantum origins, the first law of thermodynamics, $ dE = TdS $, remains a classical expression. It is anticipated that a full microscopic theory of quantum gravity will be able to reproduce this relationship, although such a theory has yet to be realized. The Euclidean approach to quantum gravity, however, allows us to derive this connection between temperature, entropy, and energy, with Euclidean time having a period of $ 1/T $.

A key question is the statistical interpretation of this entropy and its composition rule. The traditional Boltzmann-Gibbs (or Shannon) definition of entropy does not naturally lead to the type of entropy composition observed in black hole mergers. The standard Shannon entropy is
\begin{align}
S = k_B \sum_{i=1}^{W} p_i \ln \frac{1}{p_i},
\end{align}
where $ p_i $ is the probability of the $ i $-th state, and $ W $ is the number of possible states. As gravity introduces long-range interactions within the system, the validity of the above expression in the standard equilibrium thermodynamic picture becomes disputed. If we consider two independent systems, say $ A $ and $ B $, the joint probability is given by $ p_{ij}^{(A+B)} = p_i^A p_j^B $. Substituting this into the Boltzmann-Gibbs entropy formula yields the standard entropy composition:
\begin{align}
S(A+B) &= k_B \sum_{i=1}^{W_A} \sum_{j=1}^{W_B} p_i^A p_j^B \ln \frac{1}{p_i^A p_j^B}\nonumber\\ 
&= k_B \sum_{i=1}^{W_A} p_i^A \ln \frac{1}{p_i^A} + k_B \sum_{j=1}^{W_B} p_j^B \ln \frac{1}{p_j^B}\nonumber\\ 
&= S(A) + S(B).
\end{align}
This assumes that the systems are independent and that the total probability is unity. While the second assumption appears natural, the first assumption breaks down when interactions between the systems are present. In such cases, an additional term of the form $ \sqrt{S(A) S(B)} $ is expected to arise due to the interaction between the systems. To address this, one could define a correlation between the probabilities, such that $ p_{ij}^{(A+B)} \neq p_i^A p_j^B $, leading to a conditional probability $ p_i(A|B) $, which is the probability of system $ A $ given system $ B $. This Bayesian approach, however, requires a microscopic description of gravity, which is not yet available.

A solution to this problem is to generalize the definition of Boltzmann-Gibbs entropy by introducing an additional parameter. This generalization often leads to a non-additive composition rule. The first meaningful extension was proposed by R\'enyi \cite{renyi1961measures}, who introduced a parameter $ q $ and generalized the Shannon entropy as,
\begin{align}
S_q^R = \frac{k_B}{1-q} \ln \left( \sum_{i=1}^{W} p_i^q \right).
\end{align}
This generalization preserves the core ideas of Shannon's entropy while relaxing some of its postulates, such as the strong recursive postulate. The R\'enyi entropy satisfies the weak additivity relation for independent systems (see \cite{renyi1961measures} for details). The parameter $ q $ is interpreted as a measure of the distribution order, and when $ q = 1 $, the definition reduces to the Shannon entropy. Physically, $ q $ can be interpreted as the ratio of equilibrium to non-equilibrium temperature in a quasi-static, non-isothermal process \cite{e24081080}.

Later, Tsallis extended this idea further, proposing a more generalized version of entropy that produces a non-additive composition rule \cite{Tsallis1988}. Building on R\'enyi’s work, Tsallis entropy takes the form,
\begin{align}
S_q^T = k_B \frac{1 - \sum_{i=1}^{W} p_i^q}{q-1}.
\end{align}
This definition is only additive in the limit $ q \to 1 $, making it somewhat unconventional in the context of Fadeev's postulates. Tsallis and related entropies require a generalization of Fadeev's postulates, which was later proposed and established \cite{1512434}. The composition rule has the form,
\begin{align}
S_q^T(A+B) = S_q^T(A) + S_q^T(B) + (1-q) S_q^T(A) S_q^T(B).
\end{align}
The additional term $ (1-q) S_q^T(A) S_q^T(B) $ vanishes when $ q \to 1 $, reducing this to the Shannon entropy. Therefore, this entropy is non-additive. While $ q $ is often regarded as a non-additive parameter, it can also represent non-extensiveness \cite{TOUCHETTE200284}.

Non-extensiveness refers to how entropy scales with the number of microstates in a system rather than how it adds up in a composite system. For an energy-extensive system, entropy is expected to be additive, and temperature remains intensive. This is crucial for the Zeroth law and the definition of temperature \cite{PhysRevE.83.061147}. However, if energy is non-additive, even if entropy remains additive, the definition of temperature is not compatible with the Zeroth law. Black hole entropy is considered extensive when the microstates are associated with the area bits of the horizon. However, it is often termed non-extensive due to the potential 3D or 3+1D scaling of energy. This presents a more complex picture than initially apparent. While the black hole’s mass is extensive, the bits of energy responsible for the horizon's global entropy are not. In other words, although mass is additive, it is not additive in the global sense when considering the horizon mechanics. In the physical process version of the first law, it is assumed that the geometry of the black hole is not significantly affected by the amount of matter that falls in. This assumption is equivalent to ignoring the quadratic terms in the Raychauduri equation, which is necessary to evaluate the surface term where perturbations observed at infinity result in changes to the energy, while those at the bifurcate Killing horizon lead to changes in the area, which must remain equal based on linearized constraints. 

Is the first law of black hole thermodynamics valid? The first law is typically written as $ 1/T = dS/dE $, where the temperature is linked to the constant surface gravity $ \kappa $, and both entropy and energy are assumed to be additive. In dynamic cases where the geometry of the black hole changes significantly, this law may not apply. For example, in black hole mergers, the equilibrium constraints may not hold, necessitating a clearer picture. The purpose of this article is to construct a thermodynamically consistent model for processes such as black hole mergers (or simply compositions without accounting for gravitational wave emissions). 

Assuming no apparent loss of energy in the composition processes, the combined entropy becomes
\begin{align}
S_f = S(M_1) + S(M_2) + 2\sqrt{S(M_1) S(M_2)}.
\end{align}
Thus, while the energy is additive, the final entropy is not, and the process is not isothermal, as the Hawking temperature of the final black hole is lower than the lowest initial temperature. This suggests the possibility of a different statistically consistent temperature for the process, distinct from the canonical Hawking temperature. 

To illustrate this, we consider a scenario in which we begin with two well-separated black holes and end with a single static black hole. The intermediate phase is complex, so we focus only on the initial and final states. The initial entropy is given by $ S_0 = S(M_1) + S(M_2) $, and the final entropy is $ S_f = S(M_1) + S(M_2) + 2\sqrt{S(M_1) S(M_2)} $. The difference between the two entropies is:
\begin{align}
\Delta S = S_f - S_0 = 2\sqrt{S(M_1) S(M_2)}.
\end{align}
As long as $ 2\sqrt{S(M_1) S(M_2)} $ remains positive, the second law of thermodynamics is not violated, and the process is physically valid. A natural statistical framework that could explain this additional term is one based on a modified version of Shannon entropy using fractal calculus. The remainder of this manuscript aims to explore the origin, physical significance, and implications of this additional term in the context of fractional entropy.

\section{Fractional Calculus and Modified Entropy}

In \cite{UBRIACO20092516}, Marcelo R. Ubriaco proposed a generalization of Shannon entropy, following the same approach that was used to extend it to Tsallis entropy. The generalization scheme is straightforward, based on expressing Shannon entropy as
\begin{align}
S/k_B = \lim_{\zeta \to -1} \frac{d}{d\zeta} \sum_{i=1} p_i^{-\zeta}.
\end{align}
Building on this, S. Abe in \cite{ABE1997326} observed that Tsallis entropy can be written as
\begin{align}
S_q^T / k_B = \lim_{\zeta \to -1} D_q^\zeta \sum_{i=1} p_i^{-\zeta},
\end{align}
where $ D_q^\zeta $ is the Jackson q-derivative, defined as,
\begin{align}
D_q^\zeta = \frac{1 - q^{\frac{d}{d \ln \zeta}}}{\zeta(1-q)},
\end{align}
with $ q^{\frac{d}{d \ln \zeta}} $ being the shift operator in the logarithmic variable.

Further generalizing this derivative within the framework of fractional calculus, we can obtain the following
\begin{align}
_a D^q_\zeta f(\zeta) &= \left(\frac{d}{d\zeta}\right)^n \left( _a D^{q-n}_\zeta f(\zeta) \right),\\
_a D^{q-n}_\zeta f(\zeta) &= \frac{1}{\Gamma(n-q)} \int_{a}^{\zeta} \frac{f(\zeta')}{(\zeta - \zeta')^{1 + q - n}} d\zeta'.
\end{align}

While the above definition may appear complex, it can be applied to define a new entropy as,
\begin{align}
S_q^U / k_B = \lim_{\zeta \to -1} \left( \frac{d}{d\zeta} \right)^n \left( _a D^{q-n}_\zeta \sum_{i=1} p_i^{-\zeta} \right),
\end{align}
where $ n \in \mathbb{N} $ and $ q < n $ are the only constraints. 

In \cite{UBRIACO20092516}, Ubriaco demonstrated, using the Gamma function, that
\begin{align}
S_q^U / k_B = \sum_{i=1} p_i (-\ln p_i)^q.
\end{align}
Although this calculation was performed for $ n = 1 $ and $ a \to -\infty $, it can be shown that the result is independent of the choice of $ n $. This independence is crucial because the only permissible value for $ q $ is $ q < n $. As we will see in the context of black hole entropy, we require $ q = 2 $. While this generalization is valid, it affects the thermodynamic stability of the system in the canonical ensemble, leading to additional constraints.\footnote{Although Ubriaco proposed the new fractal origin of entropy in \cite{UBRIACO20092516}, the physical motivation for this concept was independently introduced by Tsallis and Cirto in \cite{Tsallis2013} within the context of black hole thermodynamics. Since Ubriaco first introduced the proper origin, we shall refer to this entropy as $ S_q^U $ and call it the Ubriaco-Tsallis-Cirto entropy, or simply fractional entropy. In the literature, the term Tsallis-Cirto is commonly used to represent a version of the area law modified by an exponent, $ (S \propto A^{\delta}) $, where $ \delta \to 1 $ recovers the traditional area law. However, this differs significantly from the approach and notions we present. Therefore, adopting a distinct notation is justified in our context. The factor $ k_B $ is included as a housekeeping term to ensure dimensional consistency.}

\subsection{Proof: Fractional entropy for any $n$}

In the definition of fractional entropy using the generalized q-derivative, the factor $ n $ plays an important role, as it determines the range of $ q $. The only constraint is that $ n $ must be a natural number ($ n \in \mathbb{N} $). In the original work \cite{UBRIACO20092516}, $ n $ was fixed at 1, and the range $ 0 < q < 1 $ was considered. This was important to establish thermal stability. Here, we have,
\begin{align}
S_q^U / k_B = \lim_{\zeta \to -1} \left( \frac{d}{d\zeta} \right)^n \left( _a D^{q-n}_\zeta \sum_{i=1} p_i^{-\zeta} \right),
\end{align}
which simplifies to
\begin{align}
S_q^U / k_B = \lim_{\zeta \to -1} \left( \frac{d}{d\zeta} \right)^n \left( _a D^{q-n}_\zeta \sum_{i=1} e^{-\zeta \ln p_i} \right).
\end{align}
upon using the definition of $ _a D_{\zeta}^{q-n} $, we get
\begin{align}
S_q^U / k_B = \lim_{\zeta \to -1} \left( \frac{d}{d\zeta} \right)^n \left( \frac{1}{\Gamma(n-q)} \sum_{i=1} \int_a^{\zeta} \frac{e^{-\zeta' \ln p_i}}{(\zeta - \zeta')^{1 + q - n}} d\zeta' \right).
\end{align}
After evaluating the integral using the Gamma function and a change of variable, and setting $ a \to -\infty $, we get,
\begin{align}
\int_a^{\zeta} \frac{e^{-\zeta' \ln p_i}}{(\zeta - \zeta')^{1 + q - n}} d\zeta' = \frac{\Gamma(n-q)}{(-\ln p_i)^{n-q}} p_i^{-\zeta}.
\end{align}
This leads to,
\begin{align}
S_q^U / k_B = \lim_{\zeta \to -1} \left( \frac{d}{d\zeta} \right)^n \left( \frac{1}{\Gamma(n-q)} \sum_{i=1} \frac{\Gamma(n-q)}{(-\ln p_i)^{n-q}} p_i^{-\zeta} \right).
\end{align}
Simplifying this, we have,
\begin{align}
S_q^U / k_B = \lim_{\zeta \to -1} \left( \frac{d}{d\zeta} \right)^n \left( \sum_{i=1} \frac{p_i^{-\zeta}}{(-\ln p_i)^{n-q}} \right).
\end{align}
Since $ \zeta $ only appears in the numerator, each derivative with respect to $ \zeta $ will multiply the whole system by an additional $ -\ln p_i $. Thus, after $ n $ derivatives, the $ n $-dependence in the denominator will cancel out, leading to
\begin{align}
S_q^U / k_B = \lim_{\zeta \to -1} \sum_{i=1} p_i^{-\zeta} (-\ln p_i)^q = \sum_{i=1} p_i (-\ln p_i)^q.
\end{align}
This result is independent of $ n $, and it allows any value for $ q $, bounded by a natural number. Specifically, $ q = 1 $ recovers the Shannon entropy.

\subsection{Composition rule}
The fractional entropy defined above is not generally additive. What is particularly interesting about the composition rule is that the general composition depends on all entropies $ S_{q'}^U $ with $ q' < q $. This is quite different from the composition of Tsallis entropy, where only the cross terms of the initial entropies are added.

Consider two independent probability distributions for two systems, $ A $ and $ B $. For $ q = 1 $, the entropy reduces to Shannon entropy, and the usual additivity is recovered. Assuming independent distributions, we have $ p_{i,j}(A+B) = p_i(A) p_j(B) $, and the composition rule becomes:

\begin{align}
&S_q^U(A+B) / k_B = (-1)^q \sum_{i=1} \sum_{j=1} p_{i,j}(A+B) \left[ \ln p_{i,j}(A+B) \right]^q\nonumber\\
&= (-1)^q \sum_{i=1} \sum_{j=1} p_{i,j}(A+B) \left[ \ln p_i(A) + \ln p_j(B) \right]^q.
\end{align}

For a positive integer value of $ q $, there will be terms of the form $ \left[ \ln p_k(X) \right]^m $, where $ m $ ranges from 0 to $ q $ in steps of unity and $k\in \{i,j\}$. Thus, the composition rule introduces a sequence of fractional entropy classes. This is particularly relevant when considering $ q = 2 $, as $q=2$ (modified) and $q=1$ (Shannon) appears together in one single expression. This indicates that there can be more than one entropy associated with the system of interest.

\section{Fractional entropy for black hole merger}

Modelling the exact features of horizons in merging black holes within a statistical framework can be a complex task. By "statistical framework," we refer to the interpretation of horizon entropy in terms of some probability distribution. This assumes that the horizon has microscopic degrees of freedom, giving rise to the entropy composition that deviates from standard additive entropy. The standard laws of horizon mechanics also refer to similar construction. However, they may be completely different from what we expect here. Thus, these "horizon states" can have a very complex existence that gives rise to the extensive thermodynamics of individual horizon thermodynamics and also gives rise to the non-additive composition when considering the black hole merger. This interplay between extensive and non-additive entropy of a single system may indicate a more profound duality that warrants further investigation. However, delving into such a duality lies beyond the scope of our current discussion.

In this context, we rely solely on the black hole entropy composition and energy conservation principles, considering only the initial and final states of the black hole. Using the fractional entropy framework, we analyze this process for $ q = 2 $. Based on the definition provided earlier, for $ q = 2 $, we can express the entropy as
\begin{align}
S_2^U(A+B)/k_B &= (-1)^2\sum_{i=1}\sum_{j=1}p_{i,j}(A+B)\left[\ln p_{i,j}(A+B)\right]^2\nonumber\\
= \sum_{i=1}\sum_{j=1}p_{i,j}&(A+B)\left[\ln p_i(A)+\ln p_j(B)\right]^2\nonumber\\
= \sum_{i=1}\sum_{j=1}p_{i,j}&(A+B)\left[\ln p_i(A)\right]^2\nonumber\\
+ \sum_{i=1}\sum_{j=1}&p_{i,j}(A+B)\left[\ln p_j(B)\right]^2\nonumber\\
+ 2\sum_{i=1}\sum_{j=1}&p_{i,j}(A+B)\ln p_i(A)\ln p_j(B).
\end{align}  
By summing over the probabilities of individual systems, we obtain
\begin{align}
&S_2^U(A+B)/k_B = \sum_{i=1}p_{i}(A)\left[\ln p_i(A)\right]^2 + \sum_{j=1}p_{j}(B)\left[\ln p_j(B)\right]^2\nonumber\\
&+ 2\sum_{i=1}p_{i}(A)\ln p_i(A)\sum_{j=1}p_{j}(B)\ln p_j(B)\nonumber\\
&= S_2^U(A)/k_B + S_2^U(B)/k_B + 2S_1^U(A)S_1^U(B)/k_B^2.
\end{align}  
Thus, the composition rule finally takes the form,
\begin{align}
S_2^U(A+B) = S_2^U(A) + S_2^U(B) + 2S_1^U(A)S_1^U(B).
\end{align}  
Unlike the standard composition of Tsallis entropy, where the additional terms are cross contributions of the same entropy type, this composition introduces the Shannon entropy ($ S_1^U $) into the expression. This specific form can serve as a composition rule for black hole entropy, provided the following assumption holds
\begin{align}
S_1^U(A)S_1^U(B) = \sqrt{S_2^U(A)S_2^U(B)}
\end{align}  
or at least  
\begin{align}
S_1^U = \sqrt{S_2^U}.
\end{align}  
While this assumption may not generally hold, proving its validity would lead to a consistent composition rule for black hole entropy under energy conservation. This also indicates that the underlying fundamental degrees of freedom can simultaneously give rise to both $S_1^U$ and $S_2^U$ like entropy, suggesting a more in-depth duality.

Can we consider the systems to be independent? As gravity and thermodynamics are deeply intertwined in black hole physics, treating black hole horizons as isolated systems may seem incomplete or non-physical. Furthermore, since gravity is a long-range interaction, conventional thermodynamic frameworks typically require the inclusion of cross-terms or non-equilibrium considerations. However, the goal of this article is not to modify the standard thermodynamic framework to account for long-range interactions like gravity. Instead, we aim to provide a robust, mathematically well-defined construction of non-additive entropy as a framework to address black hole composition-like processes. The physical connection between this non-additive framework and non-equilibrium thermodynamics remains an open question. This non-additive entropy framework treats the system as non-interacting--a natural consequence of the adopted mathematical approach.

\subsection{Case of Equal Priors}

In the previous section, we arrived at the requirement that $ S_1^U = \sqrt{S_2^U} $. We shall also see the importance of $S_1^U$ in the context of Zeroth law compatibility, and it will appear innately while defining a Zeroth law compatible temperature for fractional entropy. Here, to examine the validity of $ S_1^U = \sqrt{S_2^U} $, let us consider the case where $ p_i = 1/W $, where $ W $ represents the total number of microstates (may also be treated as the dimensionality of the whole system). In this scenario, the entropies of interests are,
\begin{align}
S_1^U &= -\sum_{i=1}^{W}p_i\ln p_i = \ln W,\\
S_2^U &= \sum_{i=1}^{W}p_i(\ln p_i)^2 = (\ln W)^2 = \ln^2 W.
\end{align}
The relation $ S_1^U = \sqrt{S_2^U} $ is exactly satisfied in this case. This demonstrates that fractional entropy provides a perfect one-to-one correspondence with black hole entropy composition under these conditions.

\section{Information Fluctuation and Fractional Entropy}

Beyond the special case discussed earlier, is there a natural connection between $ S_1^U $ and $ S_2^U $? Here we establish such a connection using the concept of information fluctuation. This approach reinforces the robustness of our analysis, where $ q \to 2 $ materializes as a special case that fits naturally within this framework. It also highlights the peculiar scenario where entropy fluctuation and information fluctuation coincide.

Consider a general probability density function $ p(x) $, where $ x $ is the random variable. The standard Boltzmann-Gibbs definition of entropy is given as,
\begin{equation}
S = \left\langle \ln \frac{1}{p(x)} \right\rangle,
\end{equation}
which arises from the concept of differential entropy. Similarly, one can generalize the notion of discrete fractional entropy to continuous variables, leading to,
\begin{equation}
S_q^U = \left\langle \ln^q \frac{1}{p(x)} \right\rangle.
\end{equation} 
For $ q = 2 $, this becomes,
\begin{equation}
S_2^U = \left\langle \ln^2 \frac{1}{p(x)} \right\rangle.
\end{equation} 
Now, consider the variance of a quantity $ \mathcal{O} $, which is defined as $ \sigma^2 = \braket{\mathcal{O}^2} - \braket{\mathcal{O}}^2 $. If we take $ \ln \left[1/p(x)\right] $ (information content) as the quantity of interest, the variance becomes,
\begin{align}
\sigma^2 = \left\langle \ln^2 \frac{1}{p(x)} \right\rangle - \left\langle \ln \frac{1}{p(x)} \right\rangle^2.
\end{align}
This implies that,
\begin{align}
\sigma^2 = S_2^U - \left(S_1^U\right)^2.
\end{align}
Interestingly, this is the definition of information fluctuation complexity. While information is technically quantified as the negative of entropy, squaring the terms ensures that both quantities are effectively the same in this context. Thus, information fluctuation is identical to entropy fluctuation for $ q = 2 $. 

Remarkably, this connection between $ S_1^U $ and $ S_2^U $ arises naturally from information theory. By definition, this fluctuation is zero for a maximally disordered or ordered system. For $ S_1^U = \sqrt{S_2^U} $ to hold, this fluctuation must vanish. Therefore, we conclude that in an ideal black hole composition that we consider here, if the final state is either maximally disordered or ordered, the black hole entropy composition can be described by the composition rule for $ S_2^U $.

Moreover, since variance is always non-negative, we obtain the inequality,
\begin{equation}
S_2^U \geq \left(S_1^U\right)^2.
\end{equation}
This provides a way to test the composition rule, where deviations from it can be interpreted as measures of information fluctuations in individual systems, which will impact the second law.

\subsection{Validity of the Second Law}

The above discussion has significant implications for the second law of thermodynamics. The second law stipulates that the final system must have greater or at least equal entropy than the initial state. This requires $ S_1^U(A)S_1^U(B) > 0 $ regardless of the details. Let us examine the scenario where $ S_1^U(A)S_1^U(B) $ approaches zero. This condition implies that the information fluctuation for each black hole must be close to its entropy. If the variance of the entropy for individual black holes approximates their $S_2^U$ entropies, $ S_1^U $ can become zero, causing the second law to reach its bound. When the variance exceeds $ S_2^U $, the second law is violated. In this context, we conjecture that, \textit{the information fluctuation of an individual black hole's entropy in a composition cannot exceed the entropy of the black hole involved.} Given
\begin{align}
S_1^U = \sqrt{S_2^U - \sigma^2},
\end{align}
if $ S_2^U = \sigma^2 $, then $ S_1^U = 0 $, and the second law is saturated. When $ S_2^U > \sigma^2 \geq 0 $, the second law holds. However, when $ 0 \leq S_2^U < \sigma^2 $, then $ S_1^U $ becomes purely imaginary. If both black holes have purely imaginary $ S_1^U $, the second law will be violated. Thus, we conclude that the violation of the second law would require an imaginary $ S_1^U $, which is not physically meaningful. Therefore, in the framework of fractional entropy, the second law remains always intact.

This, however, is a weak bound, as it does not account for energy emissions or spin components. A more stringent bound can be obtained by numerically incorporating details such as spin and orbital angular momentum. Next, we explore the significance of $ S_1^U $ under the zeroth law.

\section{Zeroth Law}

If the underlying entropy adheres to a non-additive composition rule, the definition of a Zeroth-law-compatible temperature differs subtly from the temperature derived from the Hawking temperature—the equilibrium interpretation of the first law. In \cite{PhysRevE.83.061147}, the authors demonstrate that a Zeroth-law-compatible temperature necessitates an additive functional form for entropy and energy. Within this framework, $S_1^U$ can generate a temperature consistent with the Zeroth law under fractional entropy composition. Thus, once again, $S_1^U$ naturally integrates into the definition of black hole composition, even though we started with $S_2^U$. Let us see why that is the case!

In standard thermodynamics, entropy and energy are inherently additive, and the Zeroth law is grounded in the transitivity of equilibrium states. However, in the scenario outlined above, the Hawking temperature cannot represent this empirical temperature, as black holes involved in the merger (composition) will possess a distinct Hawking temperature, and the resulting remnant exhibits a Hawking temperature lower than the minimum of the initial pair. This makes the merger a non-isothermal process in the context of conventional horizon thermodynamics. Here, we model this phenomenon through a non-additive entropy perspective, incorporating a modified definition of empirical temperature\footnote{Although this new temperature can be termed empirical, its physical process of origin, akin to the Hawking temperature, remains unknown. Here, "empirical" refers explicitly to the Zeroth-law-compatible definition as outlined in \cite{PhysRevE.83.061147}.}. A critical aspect of transitivity is that the empirical temperature reflects an intrinsic property of the system rather than a feature arising from interactions between the systems. In this context, temperature can be defined as the derivative of entropy with respect to energy, though this derivative need not be restricted to the first order. 

For transitivity to hold, additivity is a sufficient condition. Strict additivity is not required in systems governed by non-additive entropy or energy; instead, their functional forms must exhibit additivity. Specifically, even if $S_f \neq S_1 + S_2$, it suffices if $f(S_f) = f(S_1) + f(S_2)$. The same principle applies to energy, though the functional form may differ. Transitivity itself does not demand strict additivity, but for three systems with distinct functional compositions that are additive, their first derivatives with respect to the relevant physical quantity must coincide. This equivalence of first derivatives is both a necessary and sufficient condition for the validity of the Zeroth law.

\subsection{Tsallis and R\'enyi}
For instance, the standard Tsallis entropy is non-additive, with its composition expressed as,
\begin{equation}  
S_f^T = S_1^T + S_2^T + (1-q)S_1^T S_2^T.  
\end{equation} 
However, the formal logarithm of $ S $ exhibits additivity. To demonstrate this, let us multiply both sides by $ (1-q) $ and then add 1,
\begin{align}  
(1-q)S_f^T &= (1-q)S_1^T + (1-q)S_2^T + (1-q)^2 S_1^T S_2^T \nonumber \\  
1 + (1-q)S_f^T &= 1 + (1-q)S_1^T + (1-q)S_2^T + (1-q)^2 S_1^T S_2^T \nonumber \\  
1 + (1-q)S_f^T &= \left[1 + (1-q)S_1^T\right]\left[1 + (1-q)S_2^T\right]. \nonumber  
\end{align}  
Taking the natural logarithm on both sides yields,
\begin{align}  
\ln\left[1 + (1-q)S_f^T\right] =& \ln\left[1 + (1-q)S_1^T\right]\nonumber\\ &+ \ln\left[1 + (1-q)S_2^T\right] 
\end{align}  
Thus, a functional form of Tsallis entropy can be rendered additive. This highlights an important insight that a single system may possess more than one valid entropy form and associated temperature. Interestingly, each term in the above expression corresponds to the R\'enyi entropy of the system. Therefore, R\'enyi entropy provides a basis for defining an empirical temperature compatible with the Zeroth law for non-additive Tsallis entropy. To explicitly see this, we divide the above expression by $(1-q)$.
\begin{align}
\frac{1}{1-q} \ln\left[1 + (1-q)S_f^T\right] &= \frac{1}{1-q} \ln\left[1 + (1-q)S_1^T\right] \nonumber \\  
&\quad + \frac{1}{1-q} \ln\left[1 + (1-q)S_2^T\right]. \nonumber
\end{align}
Here, $\frac{1}{1-q} \ln\left[1 + (1-q)S_f^T\right]$ is the functional form of R\'enyi entropy. Once again, this observation underscores that multiple entropy definitions, arising from distinct origins, can coexist for the same system. This remains true even without detailed information about the underlying microscopic framework. Another significant feature of R\'enyi entropy is that, under the condition of equal probabilities (a priori), both R\'enyi and Shannon entropies converge to the same form, whereas Tsallis entropy remains dependent on the additional parameter. This suggests a similar relationship may hold for other entropy measures and their corresponding counterparts.

\subsection{$S_2^U$ and $S_1^U$}
For fractional entropy, the composition is given by,
\begin{align}  
S_2^U(A+B) = S_2^U(A) + S_2^U(B) + 2S_1^U(A)S_1^U(B).  
\end{align}  
When the information fluctuation is zero, this simplifies to
\begin{align}  
S_2^U(A+B) = S_2^U(A) + S_2^U(B) + 2\sqrt{S_2^U(A)S_2^U(B)}.  
\end{align}  
It follows that,
\begin{equation}  
\sqrt{S_2^U(A+B)} = \sqrt{S_2^U(A)} + \sqrt{S_2^U(B)}.  
\end{equation}  
This implies the existence of an entropy, $\sqrt{S_2^U}$, which is simply additive. Consequently, when information fluctuation vanishes, this corresponds to the $S_1^U$. Such that we have, 
\begin{equation}  
S_1^U(A+B) = S_1^U(A) + S_1^U(B).  
\end{equation}  
One can now appreciate the significance of $ S_1^U $ and the pivotal role of its connection to the information fluctuation $(\sigma^2)$. Without referencing $\sigma^2$, establishing this additive functional form would be considerably challenging.

The empirical temperature can then be defined as the derivative of this additive entropy with respect to additive energy. By definition
\begin{equation}  
\frac{1}{T} = \frac{\text{d}f(S)}{\text{d}g(E)},  
\end{equation}  
where both $f$ and $g$ are the additive functions. Here, $f$ takes the square root while $g$ takes the normal addition. Specifically, the empirical temperature becomes,
\begin{equation}  
\frac{1}{T} = \frac{dS_1^U}{dM}.  
\end{equation}  
Taking $S_1^U = \sqrt{S_2^U} \propto M$, and restoring all physical constants, we find that,
\begin{equation}  
T = \frac{1}{2\sqrt{\pi}} \sqrt{\frac{c^5 \hbar}{Gk_B^2}}.  
\end{equation}  
The key observation here is that, unlike the Hawking temperature, this empirical temperature is independent of the surface gravity of a specific horizon. Notably, this value corresponds to $\frac{1}{2\sqrt{\pi}}$ times the Planck temperature, $T_p$. This implies that all black hole horizons can be assigned this empirical temperature, regardless of their surface gravity. Consequently, any composition process in this framework will inherently be isothermal. 

The question arises: what is this new temperature? While independent of surface gravity, this temperature emerges from the statistical properties of horizon composition. It is inherently tied to the horizon and nothing beyond it. Thus, it must relate to a microscopic entity on the horizon governed by $S_1^U$-like entropy, which yields the Hawking temperature upon transition into $S_2^U$. This invites speculation about its connection to the Planck area, as a black hole with a Schwarzschild radius below the Planck length would exhibit this exact temperature as its Hawking temperature.

One can establish the relationship between the Hawking temperature and this temperature as,
\begin{align}  
T_{\text{Hawking}} = \frac{T}{2S_1^U}=\left(\frac{1}{4\sqrt{\pi}S_1^U}\right)T_p.
\end{align}
In conclusion, the Hawking temperature of any black hole horizon with entropy $ S_1^U $, defined by its energy and other parameters, is governed by the above relation. Consequently, the specific details of the surface gravity are inherently encoded within $ S_1^U $.

\subsection{Implications on Boltzmannian State Counting} 

If we equate the area law to the Boltzmann-Gibbs entropy, the Boltzmannian state counting yields the number of quantum states of the horizon as \cite{bekenstein1998quantumblackholesatoms}:  
\begin{equation}  
W = b^{4\pi R^2 / (\alpha \ell_p^2)},  
\end{equation}  
where $ b $ is an integer. From the area law, we derive $ \alpha = 4\ln b $, which characterizes the area spectrum. This implies that any area $ 4\pi R^2 $ must be an integer multiple of $ \alpha \ell_p^2 $, such that:  
\begin{equation}  
4\pi R^2 = \alpha \ell_p^2 \cdot n.  
\end{equation}  
Consequently,  
\begin{equation}  
\frac{4\pi R^2}{\alpha \ell_p^2} = \text{an integer}.  
\end{equation}  
This framework, however, is grounded in standard black hole thermodynamics, where fractional entropy composition is not incorporated.  

In the context of fractional entropy, we observe,
\begin{equation}  
S_2^U \propto R^2.  
\end{equation}  
This suggests that it is not $ \ln W $ that is proportional to the area, but rather its square. Therefore, we have:  
\begin{align}  
S_1^U &= \sqrt{S_2^U} = k_B \sqrt{\frac{\text{Area}}{4 \ell_p^2}}, \\  
S_1^U &= k_B \sum_i p_i \ln \frac{1}{p_i} = k_B \ln W = k_B \sqrt{\frac{\text{Area}}{\alpha \ell_p^2}}, \\  
\implies \ln W &= \sqrt{\frac{\text{Area}}{4 \ell_p^2}}.  
\end{align}  
Following the same reasoning as in \cite{bekenstein1998quantumblackholesatoms}, we find
\begin{equation}  
\sqrt{\frac{4\pi}{\alpha}} \left(\frac{R}{\ell_p}\right) = \text{an integer}.  
\end{equation}  
This result implies $ M \propto n $, where $ n $ is an integer. Thus, rather than the area being an integral multiple, the mass becomes an integral multiple. The Boltzmannian probability is then expressed as
\begin{align}  
p_i = \exp\left(-\sqrt{\frac{4\pi}{\alpha}} \left(\frac{R}{\ell_p}\right)\right) = \exp\left(-\sqrt{\frac{16\pi}{\alpha}} \left(\frac{M}{m_p}\right)\right).  
\end{align}  
By setting
\begin{align}  
\sqrt{\frac{\alpha}{16\pi}} m_p = \mathbb{M} \implies M = \mathbb{M} \cdot n,  
\end{align}  
where $ \mathbb{M} $ represents the mass of the fundamental black hole state within the fractional entropy framework. The implication of mass being an integral multiple may have profound connections to zero-point length. It is essential to recognize that this mass spectrum is derived from $ S_1^U $ rather than $ S_2^U $. The distribution function deviates from the Boltzmann distribution when the statistical properties of $ S_2^U $ are considered, as will be discussed in the next section.

\section{Distribution Function}
\begin{figure*}
	\includegraphics[width=0.32\textwidth]{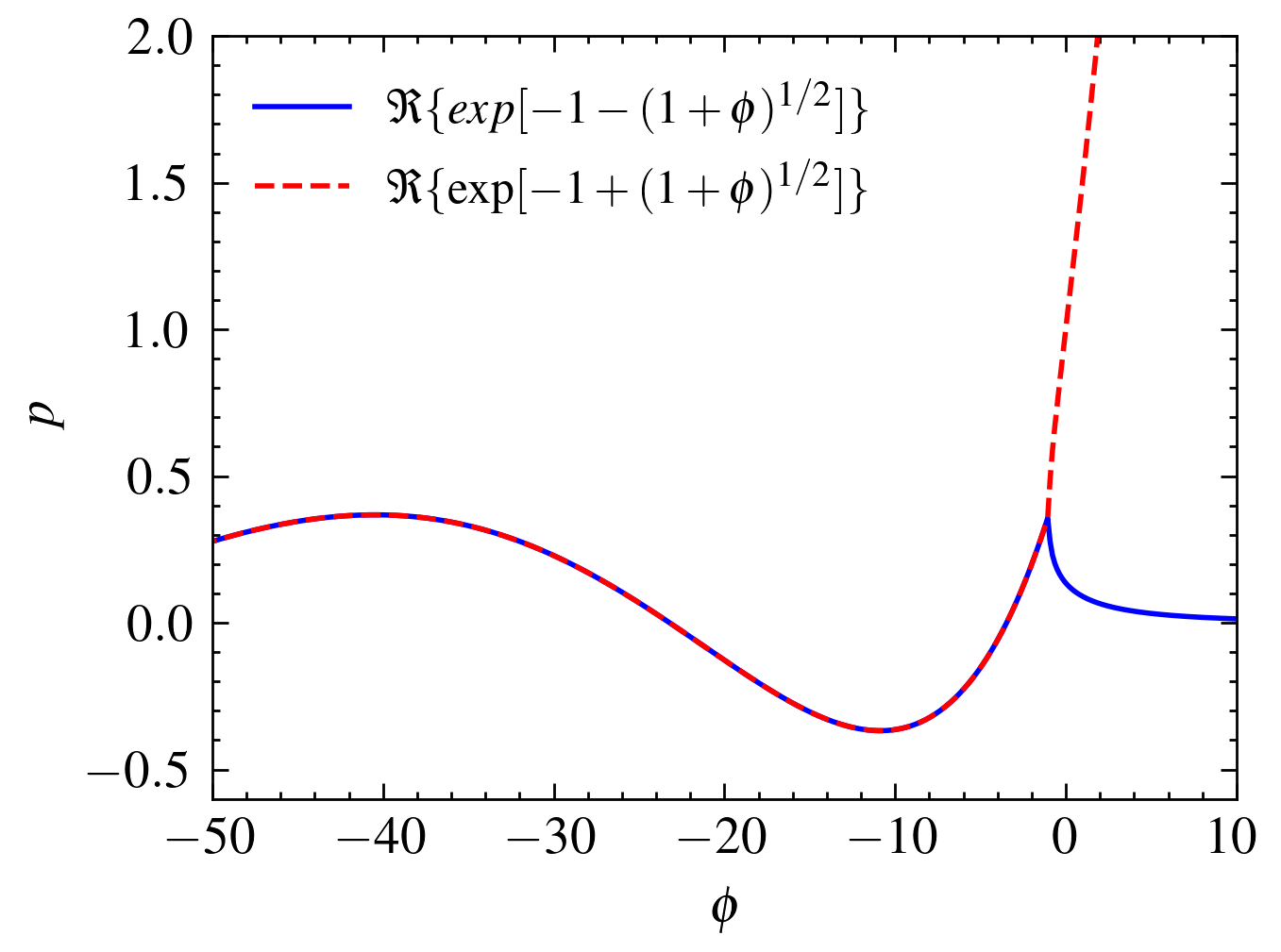}
	\includegraphics[width=0.32\textwidth]{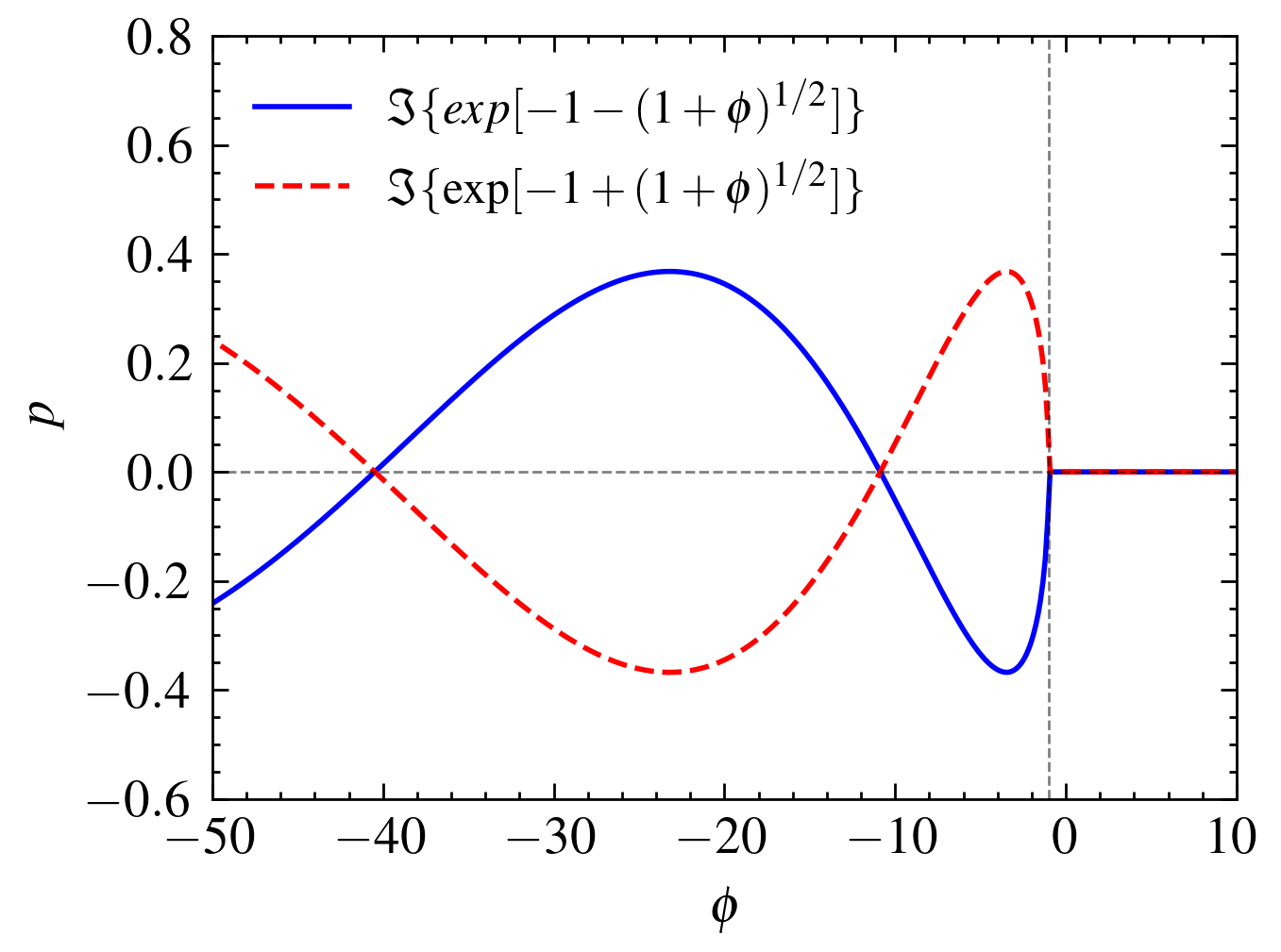}
	\includegraphics[width=0.32\textwidth]{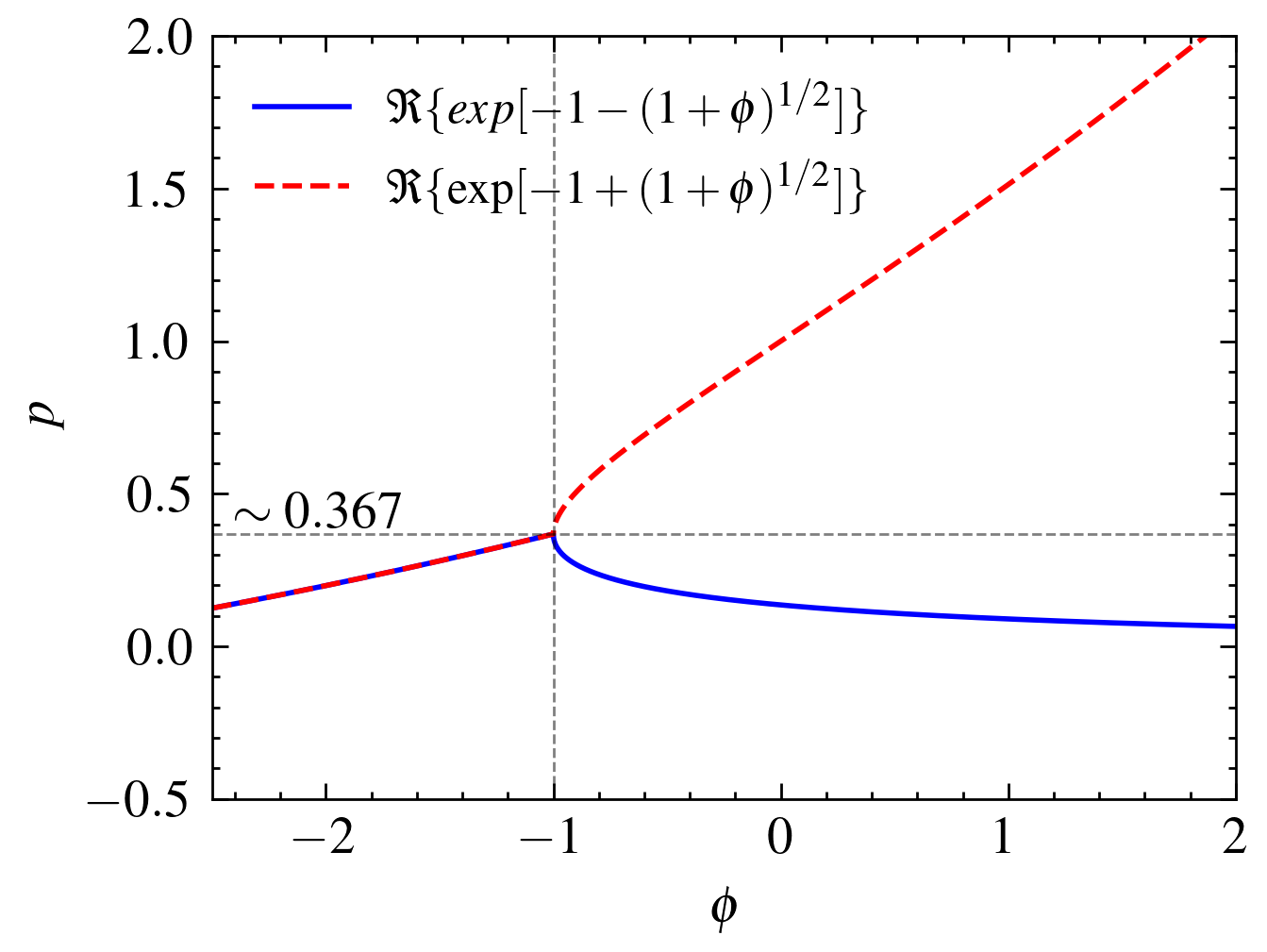}
	\caption{\label{fig:Distribution}Equilibrium probability distribution as a function of $\phi = \beta E_i + \alpha$ for $q = 2$. The first two plots depict the real and imaginary components of the distribution. For $\phi > -1$, the imaginary part becomes zero, and only one solution satisfies the distribution conditions, while the other diverges. The third plot provides a magnified view of the first plot.}
\end{figure*}
The concept of thermal equilibrium, rooted in the Zeroth law of thermodynamics, enables determining energy distribution functions by maximizing an appropriate functional. In the context of canonical ensembles, where entropy and energy are additive, the maximization problem takes the form:  
\begin{equation}  
S[p_i] - \beta \sum_{i} p_i E_i - \alpha \sum_{i} p_i = J.  
\end{equation}  
Here, $ \alpha $ and $ \beta $ are Lagrange multipliers corresponding to normalization constraints and empirical temperature, respectively. The condition for maximization, $ \frac{dJ}{dp_i} = 0 $, determines the equilibrium distribution.  

In systems governed by fractional entropy, however, the composition rules are inherently non-additive. Despite this, the additive functional form used earlier to define the empirical temperature remains applicable. Referring to \cite{PhysRevE.83.061147}, the functional can be rewritten as
\begin{align}  
f(S)[p_i] - \beta \sum_{i} p_i E_i - \alpha \sum_{i} p_i = J, \\  
S_1^U - \beta \sum_{i} p_i E_i - \alpha \sum_{i} p_i = J,  
\end{align}  
which yields results consistent with the standard framework.

However, the focus in this section is to derive the equilibrium probability distribution corresponding to $ S_2^U $. Although our immediate interest lies in the specific case $ q=2 $, a general expression for any value of $ q $ can be obtained. Starting from
\begin{align}  
\sum_{i} p_i \left(\ln \frac{1}{p_i}\right)^q - \beta \sum_{i} p_i E_i - \alpha \sum_{i} p_i = J,  
\end{align}  
and applying the Lagrange multiplier method, we arrive at the equation
\begin{align}  
\left(\ln \frac{1}{p_i}\right)^q - q\left(\ln \frac{1}{p_i}\right)^{q-1} - \beta E_i - \alpha = 0.  
\end{align}  
Substituting $ x = \ln \frac{1}{p_i} $, this simplifies to
\begin{align}  
&x^q - qx^{q-1} - \beta E_i - \alpha = 0, \\  
\implies &(x - q)x^{q-1} = \beta E_i + \alpha.  
\end{align}  

This approach provides a robust framework for exploring equilibrium distributions in fractional entropy systems, extending beyond the conventional additive entropy paradigm. In the above calculation, we assumed the entropy to be $S_2^U$ (non-additive) and energy to be additive.

When $ q = 2 $, we have the equation
\begin{align}
\ln \frac{1}{p_i} = 1 \pm \sqrt{1 + \beta E_i + \alpha}
\end{align}
Substituting $ \beta E_i + \alpha = \phi_i $, this simplifies to:
\begin{align}
p_i = \exp\left( -1 \mp \sqrt{1 + \phi_i} \right)
\end{align}
This distribution deviates from the standard equilibrium distribution. Notably, it can take complex values, which presents an interesting aspect for further study, though this lies beyond the scope of this work. However, let us delve deeper into the distribution.

Since the equilibrium distribution is complex, we must examine the region where it becomes purely real and exhibits standard properties. To simplify, we focus on $ \phi $ as the key variable. The behaviour of $ p_i $ depends on $ \phi $ and can result in complex values. Additionally, $ p_i $ has two possible values due to the square root term. A degeneracy occurs at $ \phi = -1 $, and for $ \phi $ greater or smaller than $ -1 $, the distribution bifurcates, where one solution diverges and no longer behaves like a valid probability distribution. Therefore, we choose
\begin{align}
p_i = \exp\left( -1 - \sqrt{1 + \phi_i} \right)
\end{align}
This is a physically reasonable choice. Notably, the imaginary part of $ p_i $ vanishes for $ \phi > -1 $, while for $ \phi < -1 $, it exhibits oscillatory behaviour. The oscillation frequency depends on $ \phi $ and diminishes as $ \phi \to -\infty $. Furthermore, a phase shift of $ \pi $ between the imaginary parts results from the $ \pm $ in the complete solution. In contrast, the real part remains unaffected by the $ \pm $ factor for $ \phi < -1 $. See FIG. (\ref{fig:Distribution}) for illustrations.

The normalization factor $ \alpha $ can be estimated numerically for specific cases, but there is a maximum value for the probability distribution. If the probability exceeds this limit, it is not physically plausible. In the general framework of fractional entropy, this maximum value depends on $ q $, which in our case is $ q = 2 $. For standard Shannon entropy, where $ q = 1 $, the maximum value is 1, with appropriate normalization by $ \alpha $. However, for $ q \neq 1 $, this value will always be less than the maximum for $ q = 1 $. Without normalization, we observe that for $ q = 2 $, the maximum value is approximately 0.367. This behaviour is similar to an exponential distribution, though it differs from the conventional $ q $-exponential distribution associated with Tsallis entropy. The point of maximum probability is also significant, as it shifts from $ \phi = -1 $ depending on the value of $ q $. This shift may be linked to the dimensionality of the system. In \cite{PhysRevE.88.052107}, the authors, including Tsallis, demonstrate this feature numerically. They define $ q = \frac{D}{D-1} $, where $ D $ is the system's dimensionality. With $ q = 2 $, we are considering a two-dimensional system consistent with the assumptions regarding counting surface degrees of freedom in conventional black hole mechanics. This assumption is based on the principle that for an anomalous $ D $-dimensional system, $ \ln[W(L)] \propto L^{D-1} $. Therefore, $ S_q^U $ will be extensive (i.e., it scales) when $ q = \frac{D}{D-1} $. As a result, the probability distribution can be mapped to the system's dimensionality. However, this is just one possible choice, and there is no strict requirement that the entropy of the black hole must be extensive or that the black hole is a 3+1-dimensional system.

The Boltzmann distribution, being exponential, suggests that its generalization would also exhibit similar behaviour. However, there are key differences to consider. We have the expression
\begin{align}
p_i = \exp\left( -1 - \sqrt{1 + \beta E_i + \alpha} \right)
\end{align}
For normalization, we require
\begin{align}
\sum_i p_i = \sum_i \exp\left( -1 - \sqrt{1 + \beta E_i + \alpha} \right) = 1
\end{align}
Assuming $ E $ is a continuous variable, the normalized probability distribution function is
\begin{align}
p(E) = \frac{\beta \exp\left( \sqrt{\alpha + 1} - \sqrt{1 + \beta E + \alpha} \right)}{2\left( \sqrt{\alpha + 1} + 1 \right)}
\end{align}
with the support domain $ E \in [0, \infty) $. Here, $ \sqrt{\beta} $ is expected to be a positive real number and $ \alpha \geq -1 $. This distribution differs from the standard Boltzmann distribution, representing an exponential function, which would take the form $ \beta e^{-\beta x} $ in the same domain. Thus, the distribution depends on $ \alpha $, which needs to be estimated. In the conventional Maxwell-Boltzmann distribution, these parameters are related to temperature and chemical potential. However, $ \beta $ is not the empirical thermodynamic temperature, as the entropy we are working with is not additive. The additive form of the entropy is $ \sqrt{S_2^U} $, applicable when there is no information fluctuation. Using this form will lead to the standard Boltzmann distribution, where $ \alpha $ becomes irrelevant, and $ \beta $ represents a temperature compatible with the Zeroth law. Depending on the energy distribution's characteristics, this distribution may also resemble a hyperbolic distribution with a similar form but a quadratic dependence on the random variable.

\section{Stability}

\begin{figure}[t]
	\includegraphics[width=0.45\textwidth]{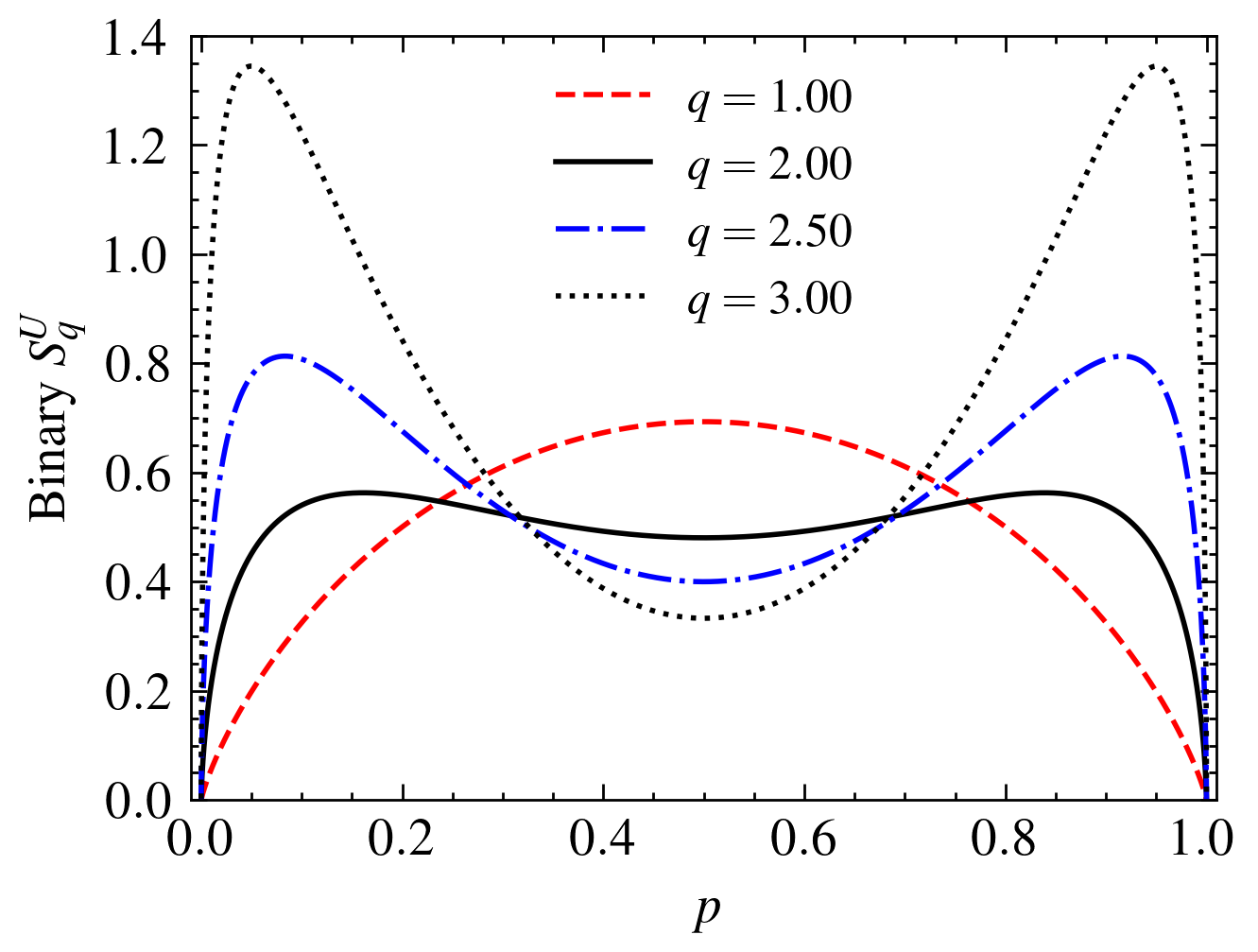}
	\caption{\label{fig:entropy}The entropy as a function of the binary probability distribution for different values of $ q $. For $ q \to 0 $, the entropy becomes independent of $ p $ and reaches unity. For $ q > 1 $, entropy is symmetric about $ p = 1/2 $, with an inverted Mexican hat profile. The maximum entropy now depends on the value of $ q $. For $ q = 2 $, the maxima occur at $ p \sim 0.16 $ and $ p \sim 0.84 $, with $ S_2^U(\text{max}) \sim 0.56 $.}
\end{figure}

We begin by examining the behaviour of the binary entropy system. When $ q = 1 $, we recover the conventional Shannon entropy, while as $ q \to 0 $, the entropy reaches unity by definition. For values of $ q > 1 $, the entropy adopts an inverted Mexican hat profile, acquiring two symmetric maxima. These maxima shift symmetrically as $ q $ increases, as demonstrated in Figure \ref{fig:entropy}, which clearly shows the shift and symmetry of the maxima with varying $ q $.

Following \cite{UBRIACO20092516}, we observe that the thermodynamic stability of the system depends on the value of $ q $ used in the entropy definition. The bound on $ q $ is a positive integer $ n $, which we have shown to be infinite, implying that $ q $ can be bounded by any positive integer. While $ q = 2 $ is the choice for black hole composition, this does not automatically imply thermal stability. To confirm stability, we must examine the second derivative of entropy with respect to the probability distribution, which must be negative to ensure concavity, which implies thermal stability.

For the standard Shannon entropy ($ q = 1 $), the entropy is concave for a two-dimensional system, as seen in Figure \ref{fig:entropy}. However, for $ q > 1 $, the concavity is not always satisfied. Specifically, the information fluctuation complexity is zero for a maximally mixed and pure state, suggesting that a two-level system may not be sufficient to account for the dynamics of black hole composition processes.
\begin{figure*}
	\includegraphics[width=0.32\textwidth]{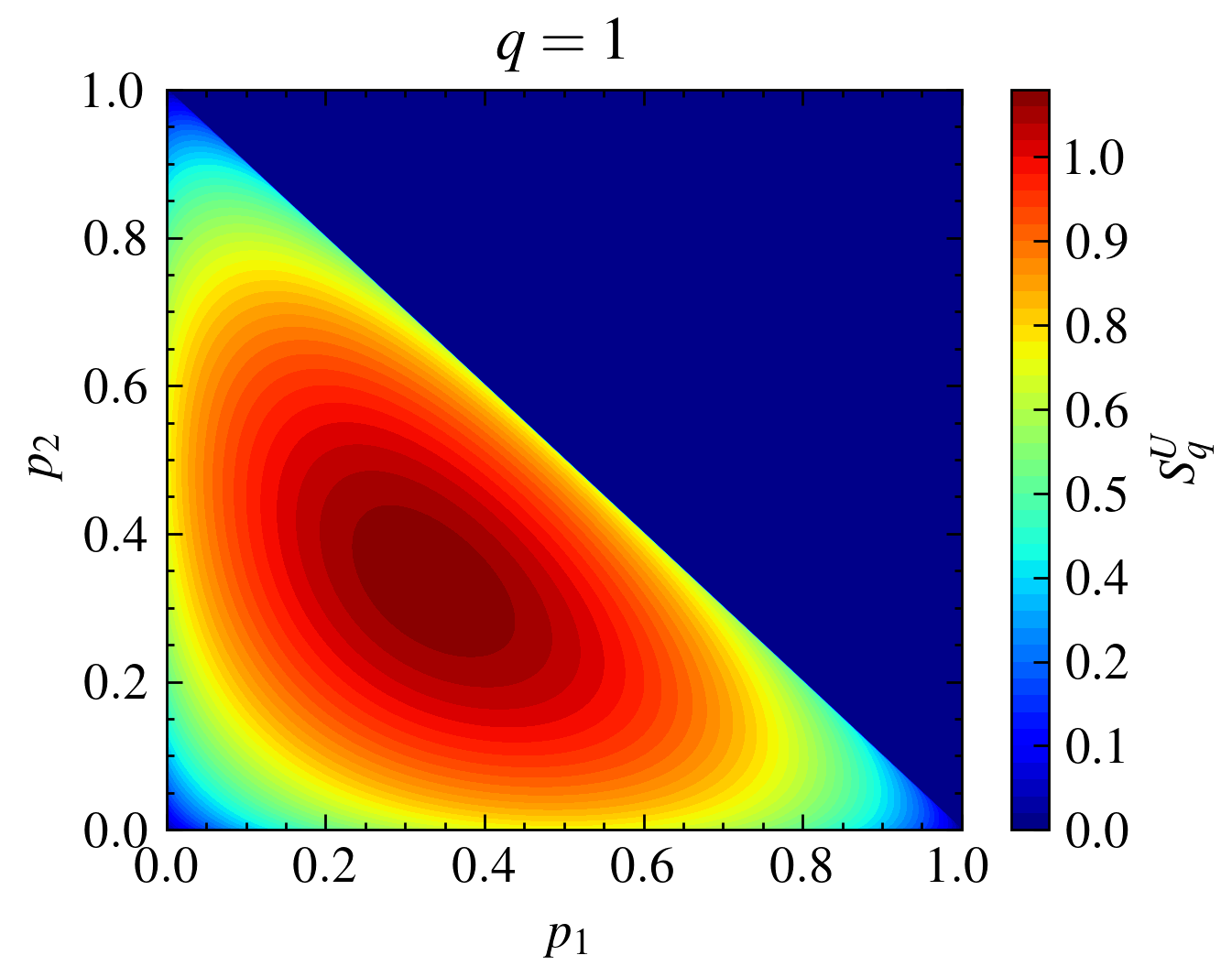}
	\includegraphics[width=0.32\textwidth]{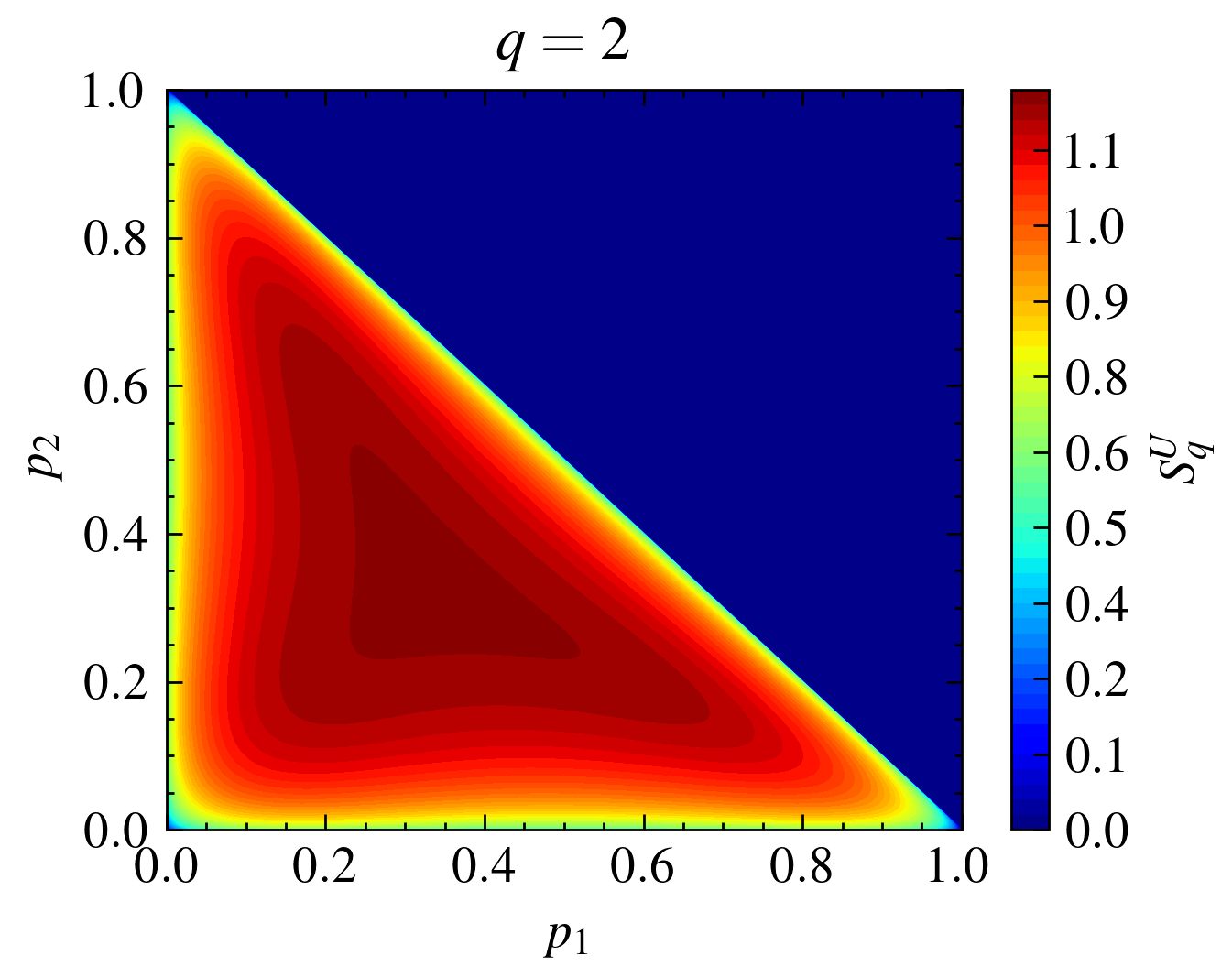}
	\includegraphics[width=0.32\textwidth]{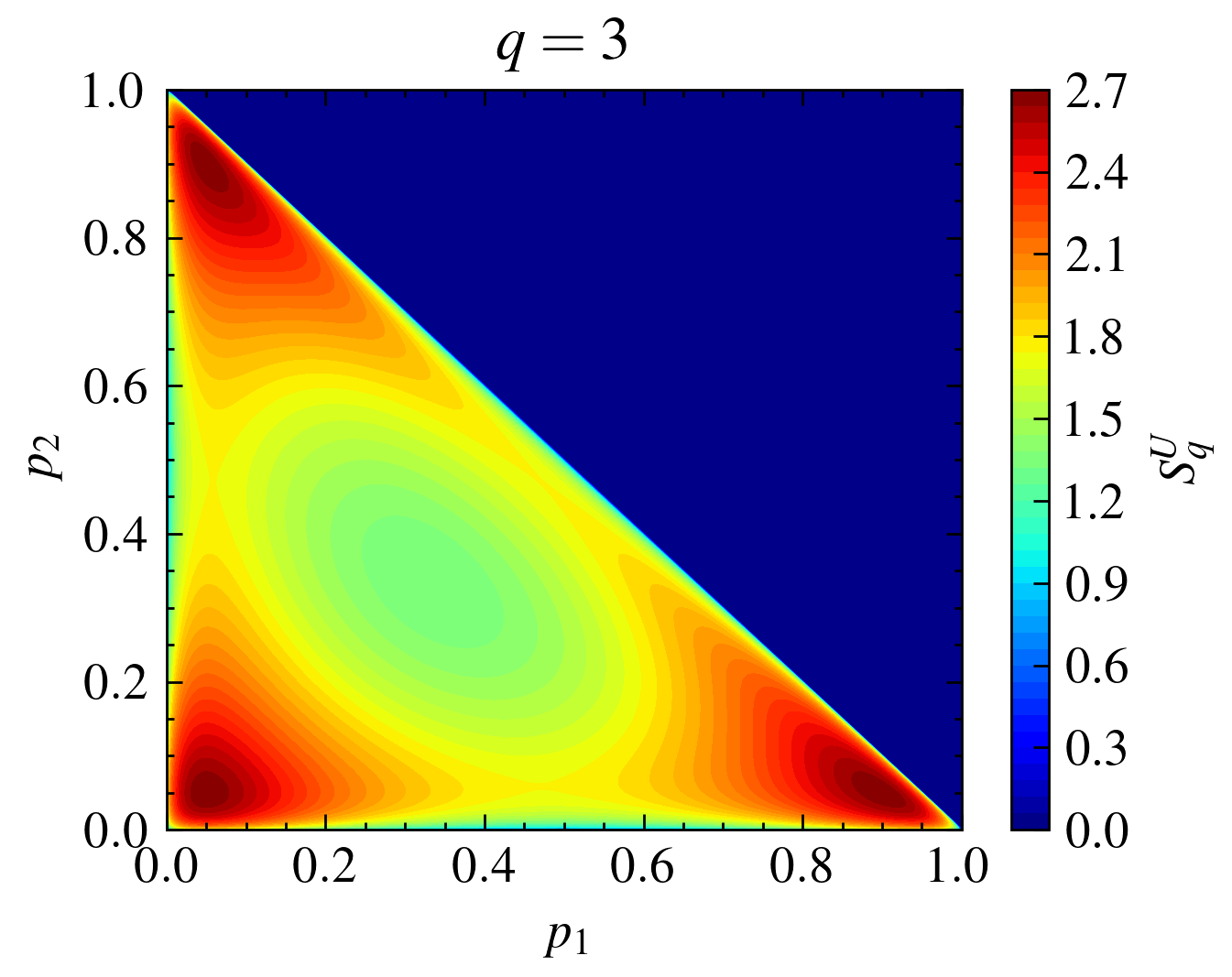}
	\caption{\label{fig:ThreeEntropy}Entropy $ S_q^U $ of a three-level system as a function of probability for different values of $ q = (1, 2, 3) $.}
\end{figure*}
\begin{figure*}
	\includegraphics[width=0.32\textwidth]{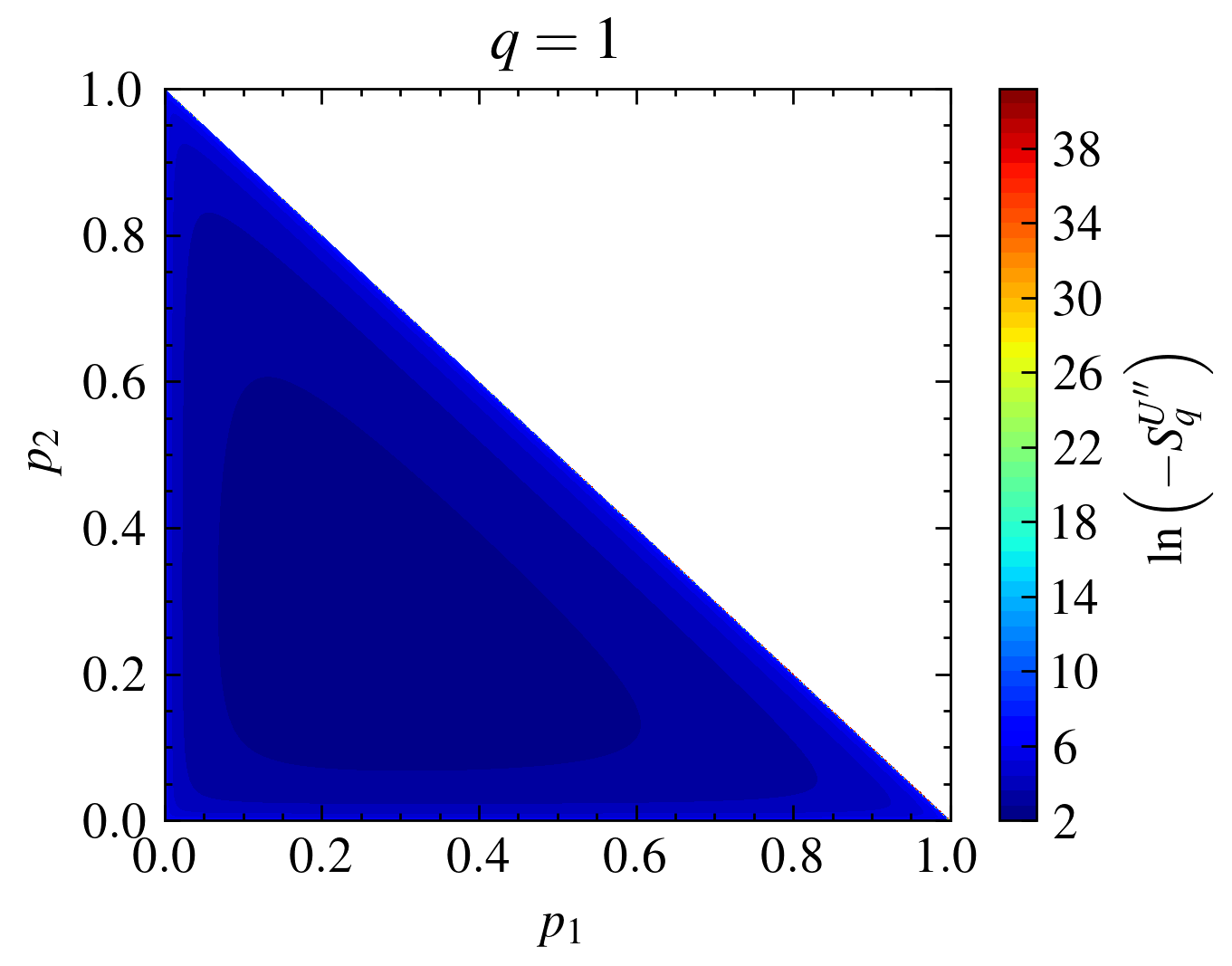}
	\includegraphics[width=0.32\textwidth]{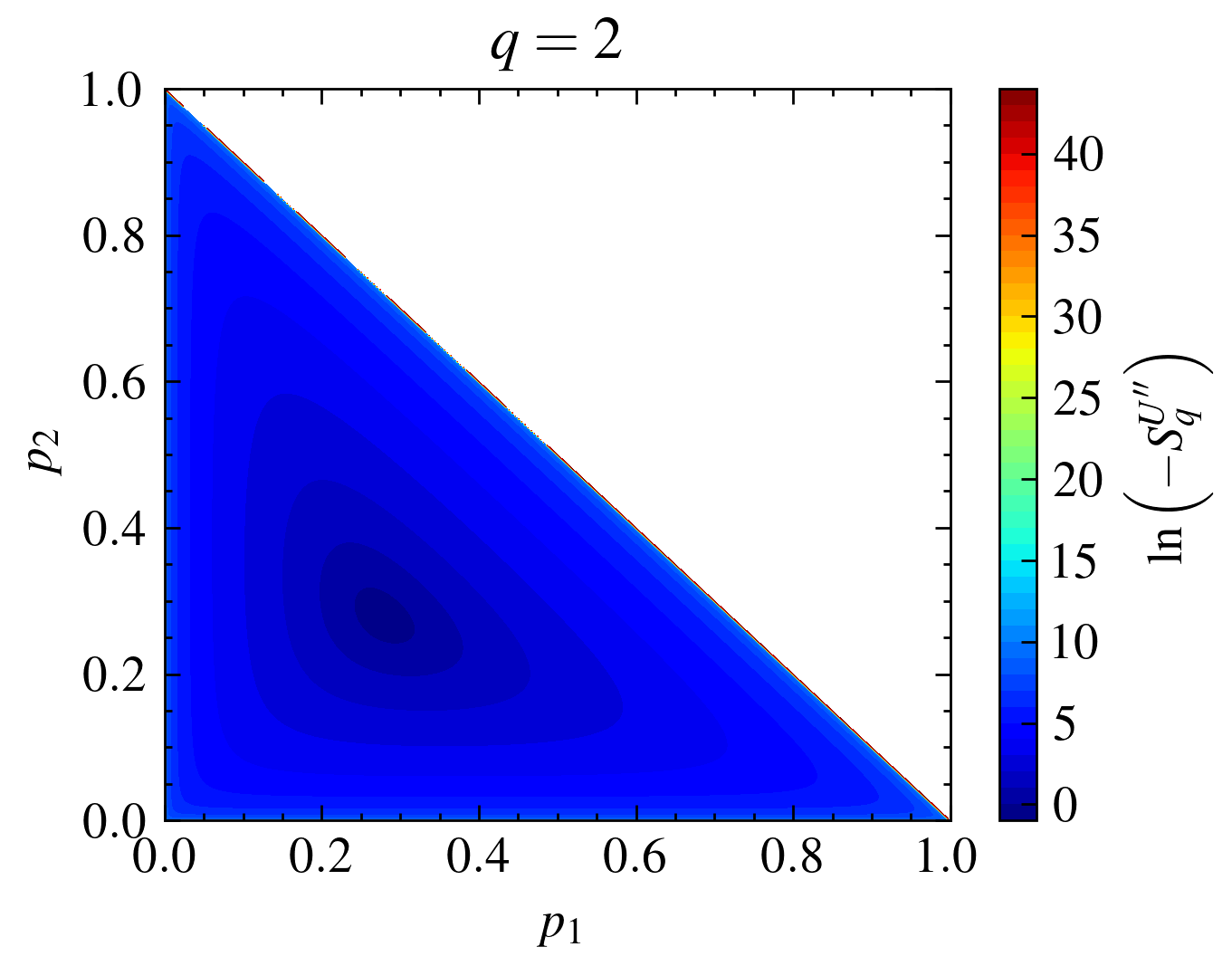}
	\includegraphics[width=0.32\textwidth]{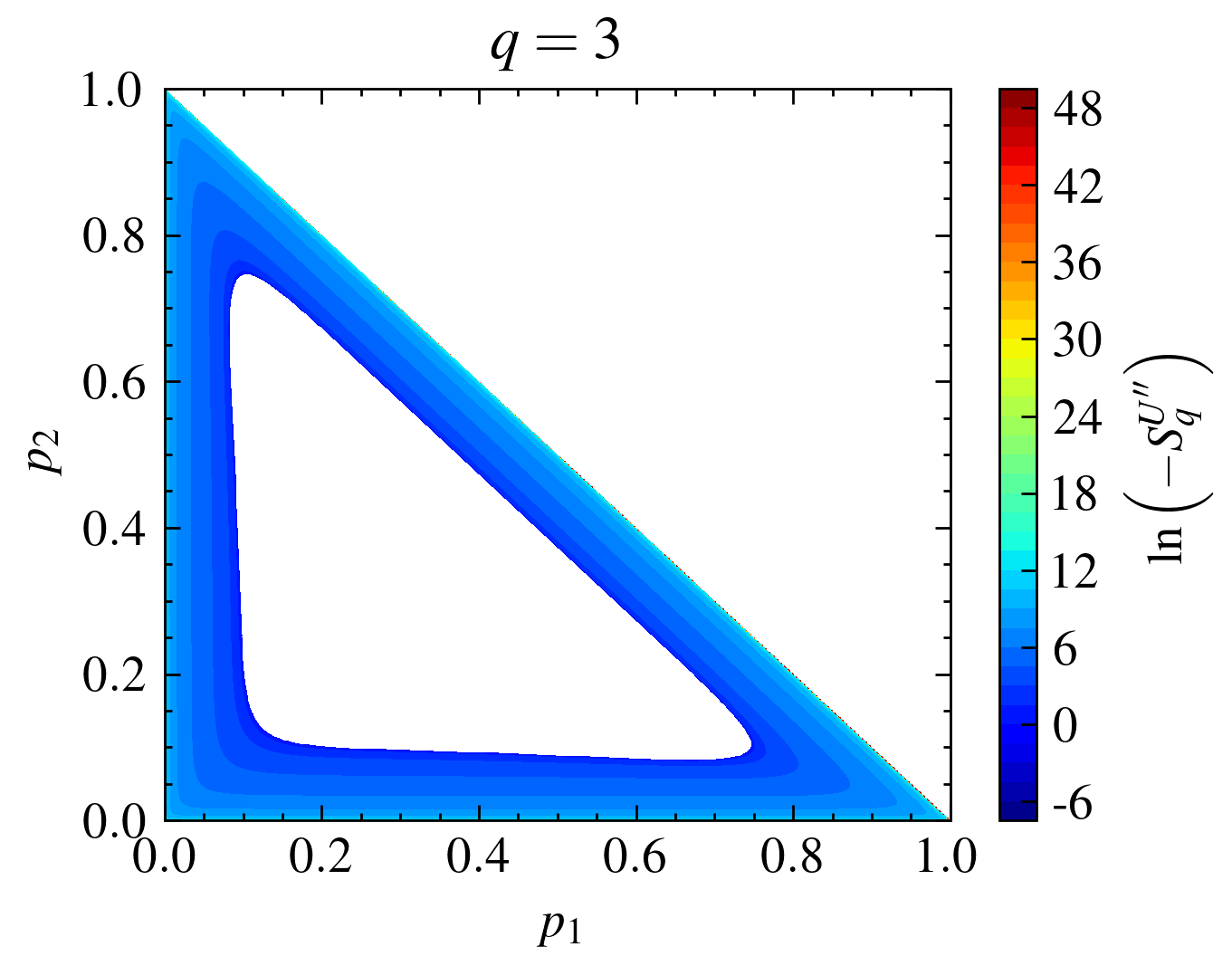}
	\caption{\label{fig:ThreeEntropy2}Logarithm of the negative second derivative of entropy $ \ln(-S_q^{U''}) $ for a three-level system as a function of probability for different values of $ q = (1, 2, 3) $. The white region indicates \texttt{NaN} (not a number) in Python indicating positive second derivative $(S_q^{U''}>0)$.}
\end{figure*}

We derive the second and first derivatives of the entropy as follows
\begin{align}
S^{U''}_q = -\frac{q}{W^2} \left[ S_{q-1}^U - (q-1) S_{q-2}^U \right]
\end{align}
\begin{align}
S^{U'}_q = \frac{q}{W} S_{q-1}
\end{align}
where $ W $ is the dimension of the system, and $ p_i = 1/W $ for simplicity. Inserting these into the general composition rule, we obtain the condition for thermal equilibrium as,
\begin{align}
\frac{1}{2} S^{U''}_q - \frac{2^q - 1}{2W^2} \left[ q S_{q-1}^U + (q - q^2) S_{q-2}^U \right] < 0
\end{align}
For the case of black hole composition, $ q = 2 $ does not always satisfy this condition. Substituting $ q = 2 $, we arrive at the inequality
\begin{align}
-\frac{4}{W^2} \left( \ln W - 1 \right) < 0
\end{align}
This condition holds only when $ \ln W - 1 > 0 $, or $ W > e $. Consequently, a two-level system is not thermally stable for $ q = 2 $, necessitating a system with more than two levels, such as a qutrit. \footnote{This numerical value will also depend on the base of the logarithm used. For simplicity, we choose it to be $e$. Otherwise, the expression will have an additional factor of $\ln_e(\text{base})$.}.

For a three-level system, entropy is positive in the region $ 0 \leq p_i \leq 1 $, but concavity is violated when $ q = 3 $. In Figure \ref{fig:ThreeEntropy2}, we plot the logarithm of the negative second derivative of entropy, where a positive second derivative results in a complex value, and Python identifies this as "not a number" leaving a space in the plots. This shows that the system remains thermally stable for $ q = 2 $ but not for $ q > 2 $.

Thus, the condition for thermal stability is given by
\begin{align}
-\frac{q}{W^2} 2^{q-1} \left[ \ln^{q-1} W - (q-1) \ln^{q-2} W \right] < 0
\end{align}
which demands that
\begin{align}
q < 1 + \ln W, \text{ for } q > 0 \text{ and } W > 1
\end{align}
\begin{align}
W > \exp(-1 + q), \text{ for } q > 1 \text{ and } W > 1
\end{align}
Hence, for $ q = 2 $, we require more than a two-level system, making qubits or bits unsuitable for information carriers in our context.

Finally, we examine the behaviour of the system for different values of $ W $ and $ q $, expressed by the function
\begin{align}
X = -\frac{q}{W^2} 2^{q-1} \left[ \ln^{q-1} W - (q-1) \ln^{q-2} W \right]
\end{align}

\begin{figure}
	\includegraphics[width=0.45\textwidth]{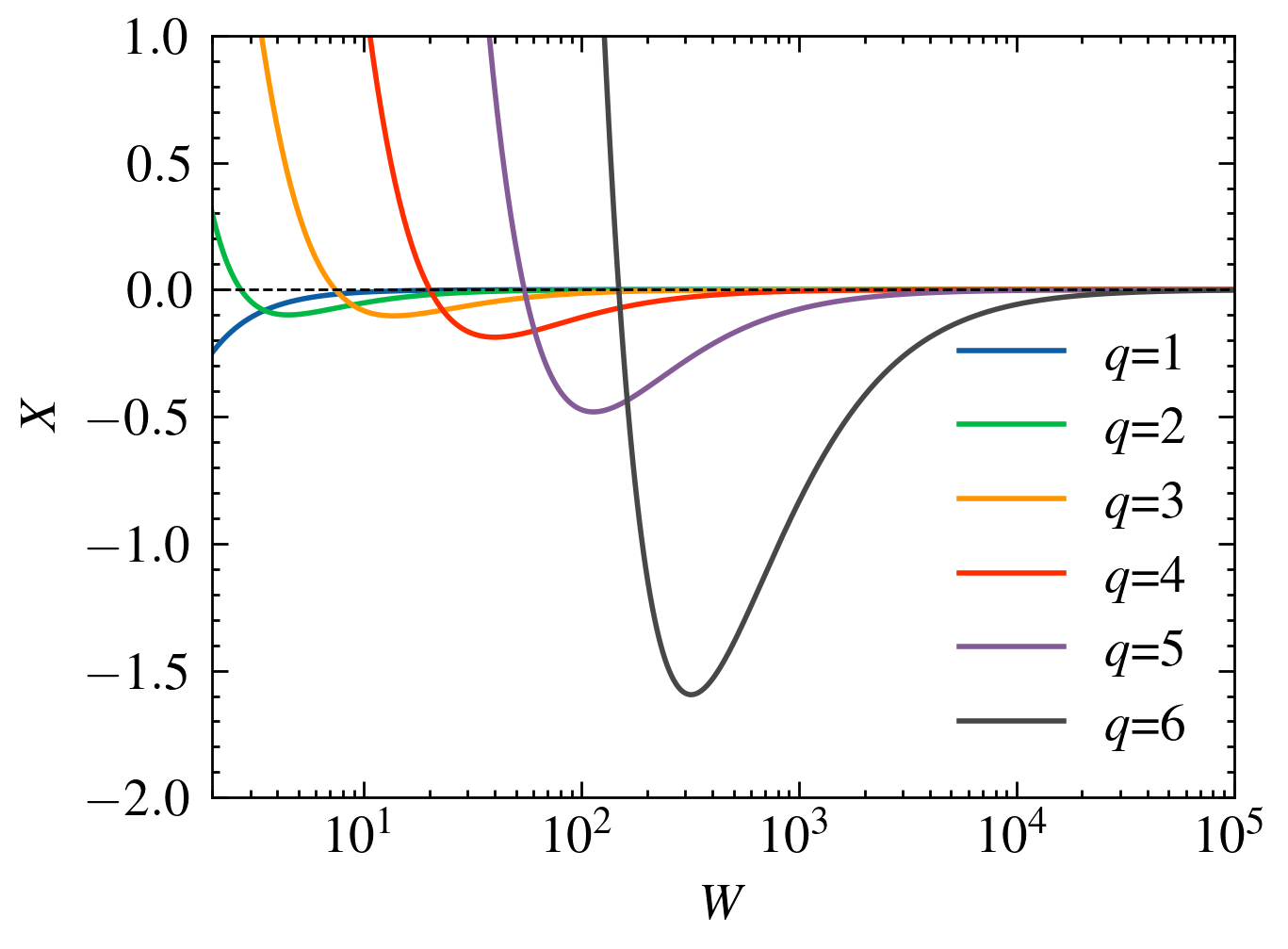}
	\caption{\label{fig:c2} Stability criterion $ X $ as a function of $ W $ for different values of $ q $. The system is considered thermally stable when $ X < 0 $ and thermally unstable when $ X > 0 $. (Colour online)}
\end{figure}

\begin{figure}
	\includegraphics[width=0.45\textwidth]{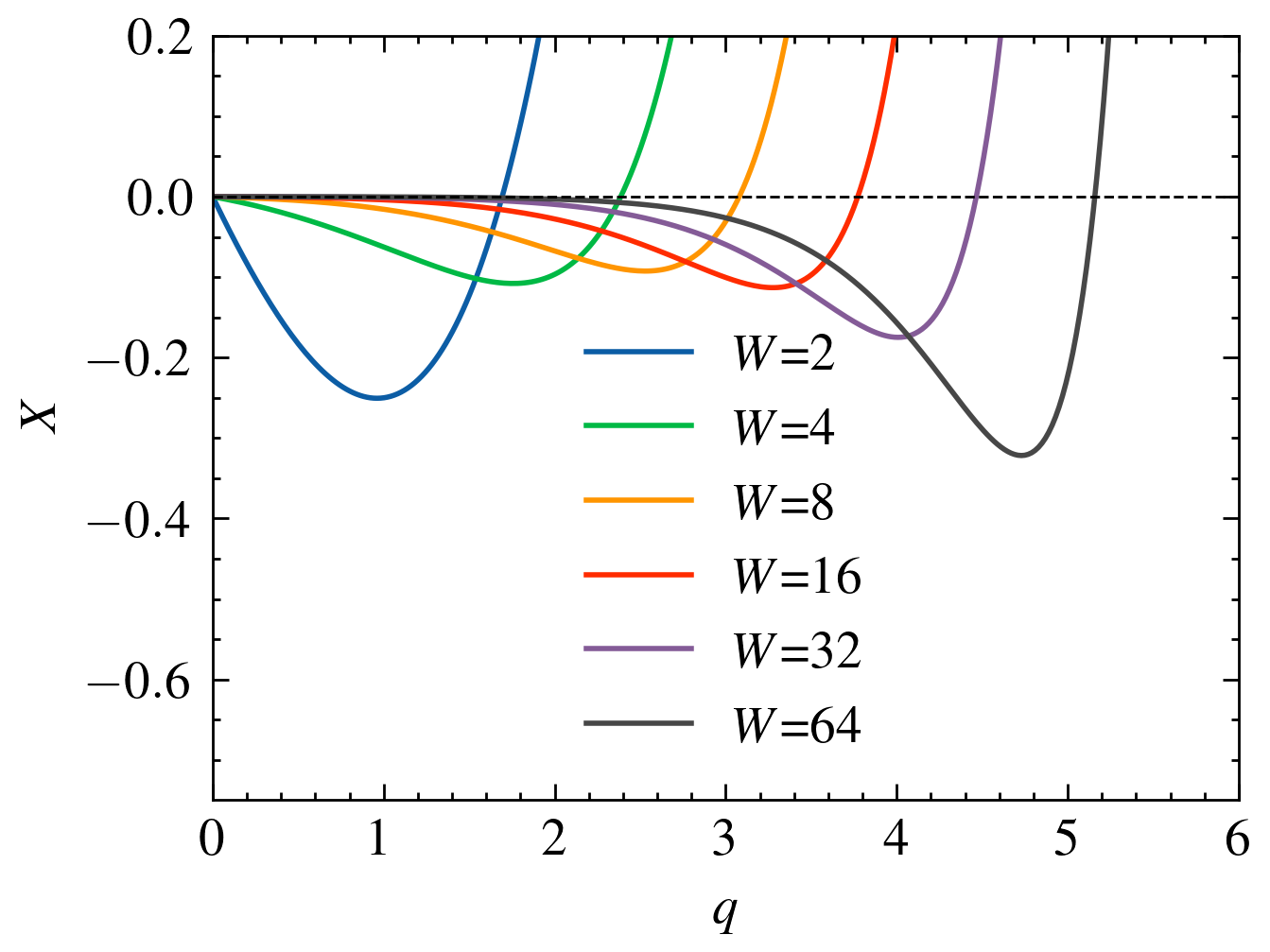}
	\caption{\label{fig:c1} Stability criterion $ X $ as a function of $ q $ for different values of $ W $. The system is considered thermally stable when $ X < 0 $ and thermally unstable when $ X > 0 $. (Colour online)}
\end{figure}

In conclusion, thermal stability is achieved when the quantity $ X $ is negative, though this condition is not universally satisfied for all values of $ W $ and $ q $. As $ W \to \infty $, the system's thermal stability becomes independent of $ q $, as seen in FIG. (\ref{fig:c2}). For $ q=1 $, thermal stability is guaranteed for all values of $ W $. For the purposes of this discussion, we assume that $ W > 1 $, which allows for fractional values in a mathematical sense, though their physical implications require further investigation.

For different values of $ q $, a minimum threshold exists for $ W $ to ensure thermal stability. This was illustrated for two- and three-level systems earlier. For example, in FIG. (\ref{fig:c1}), we observe that for $ W=2 $, the system is stable only for $ q < 2 $, and for $ W=64 $, stability is maintained up to $ q < 6 $. Thus, the thermal stability of the system is significantly influenced by the specific values of $ W $ and $ q $. Notably, for $ q=2 $, a two-level system (with a natural logarithm base $ e $) is insufficient to guarantee thermal stability, which may explain why studies involving $ q > 1 $ are relatively rare in the literature.

\section{Conclusion}

In the manuscript, we focus explicitly on the entropy composition of black holes under energy conservation in an idealized situation and study it under the lens of fractional entropy. After setting the stage, we introduced the features of fractional entropy that can reproduce the exact composition rule. From this, we explored the details of the class of fractional entropy along with its composition rule and see that their definition for $q=2$ itself invokes $S_2^U$ and $S_1^U$ in the composition rather than $S_2^U$ alone as in general Tsallis or R\'enyi parametrizations. For any values of $q'$, the whole spectrum of $S_{q\leq q'}^U$ appears in the composition, making the composition rather unique. Interestingly, we see that for $q=2$, this connection between $S_2^U$ and $S_1^U$ can be made by invoking the notion of information fluctuation complexity. This turns out to be an advantage in putting stronger implications on the validity of the second law. We also explored the mathematical details of the fraction entropy and found that the value of $q$ is bounded by any natural number rather than unity, as illustrated in the literature. However, this comes with the prize that affects the thermal stability, which is always valid for $0<q\leq 1$. Upon extending this bound, we show that the thermal stability can be observed upon increasing the system's dimension ($W$). We see that the system becomes thermally stable for very large $W$ irrespective of the value of $q$. However, for smaller $W$, the thermal stability is contingent on $q$ and vice versa. 

Besides the implications on stability and the second law, we studied the Zeroth law compatible empirical temperature associated with the fraction entropy. We found that the same $S_1^U$ is additive by definition, and it gives the empirical temperature of the system. This empirical temperature turns out to be a universal constant, which could have strong implications for the theories or phenomenologies of quantum gravity. We also establish the connection between this new empirical temperature and the Hawking temperature associated with the particular horizon based on the $S_1^U$ defined. Thus, establishing the transformation that depends on the specific $S_1^U$, we see that $S_1^U$ itself encodes the details of the surface gravity. In the whole analysis, we assume that $S_2^U$ is the one that is responsible for the area law. This empirical temperature allows us to picture the black hole composition as an isothermal process, although the physical picture with Hawking temperature indicates otherwise.

Furthermore, we explored the traditional Boltzmannian state counting in the context of standard Shannon and fractional entropy and saw that when the Shannon entropy generates a spectrum for area, the fraction entropy generates an equidistant spectrum for mass. We then explicitly explored the thermal equilibrium distribution based on this fractional entropy and studied the features of the equilibrium distributions. The solutions indicate a wide spectrum, including complex distributions. As future scope, it would be interesting to explore the complex region of the probability distribution in the context of positive real representation as illustrated in \cite{10.1063/1.531906}. Here, we explored only the one that showed convergence in the energy distribution. However, the deviation from Boltzmann distribution and the connection between thermal stability with a lower limit on dimensionality shows the need for a modified Boltzmannian state counting to understand the details of horizon degrees of freedom. Thus, we establish that the traditional bits or qubits may not be the ideal candidates in the picture we follow with fractional entropy, although it may be suitable in conventional thermodynamics where the area law is proportional to the Shannon entropy.

This work opens several new areas that need further exploration, which are indeed interdisciplinary. We need to understand the features of the equilibrium distribution obtained, as it is significantly different from the standard Boltzmann or Hyperbolic distributions. We need to see if there are general connections between fractional entropy, information fluctuations, and thermodynamic fluctuations \cite{PhysRevD.105.043534}. We need to study further the physical origin of the empirical temperature derived and its implications in area quantization and zero point length. These could have a profound impact on solving the information paradox. By the end of this work, we came across the work in \cite{volovik2024tsalliscirtoentropyblackhole}, which has indicated a similar result considering quantum tunnelling in black holes. These results suggest that the area spectrum is quadratic while the mass has an equidistant spectrum. Although the results agree, our approach and that of what was discussed in \cite{volovik2024tsalliscirtoentropyblackhole} are different.

Horizon thermodynamics has played a pivotal role in shaping the construction of dark energy within holographic frameworks and modified cosmology in general. As proposed in this work, understanding it through the lens of fractional entropy offers a distinct perspective compared to the approaches outlined in \cite{Manoharan2023, Manoharan2024}. This alternative framework holds the potential to significantly advance our understanding of the emergent nature of gravity and space from a statistical standpoint.

\begin{acknowledgments}
The authors express their sincere gratitude to Titus K. Mathew for discussions on non-additive entropy in the context of cosmology, which inspired the need for this work. They also extend their heartfelt thanks to Navaneeth Krishnan for insightful discussions on Shannon entropy and to M. Dheepika for her comments regarding Tsallis entropy. This research was supported by the CSIR-NET JRF/SRF program, Government of India, under Grant No. 09/239(0558)/2019-EMR-I. The authors appreciate Rhine Kumar A. K., Vishnu A. Pai, and Vishnu S. Namboothiri for their thoughtful feedback.\\
\end{acknowledgments}

\paragraph*{\textbf{Data Availability Statement:}} This manuscript has no associated data or the data will not be deposited.\\

\paragraph*{\textbf{Use of AI tools:}} The authors acknowledge the assistance of ChatGPT and Grammarly in improving the grammar and language usage in this article. After utilizing these tools, the authors thoroughly reviewed and edited the entire content, taking full responsibility for its accuracy and quality.\\

\paragraph*{\textbf{Python modules used:}} The following Python packages were used for the numerical calculations and illustrations presented in this manuscript: NumPy \cite{harris2020array}, SciPy \cite{2020SciPy-NMeth}, SciencePlots \cite{SciencePlots}, and Matplotlib \cite{Hunter:2007}.

\end{document}